\documentclass[12pt]{ronbun}

\usepackage{amsmath,amssymb}
\usepackage{cite}
\usepackage{bm}
\usepackage{dcolumn}

\newcommand{\nn}{\nonumber}
\newcommand{\e}{{\rm e}}
\newcommand{\tr}{\rm Tr}
\newcommand{\del}{\delta}
\newcommand{\ra}{\rangle}
\newcommand{\la}{\langle}

\newcommand{\al}{\alpha}

\renewcommand{\th}{\theta}

\newcommand{\s}{\scriptscriptstyle}

\numberwithin{equation}{section}

\begin{document}

\begin{flushright}
\parbox{4.2cm}
{KEK-TH-891 \hfill \\
%KUNS-XXXX \hfill \\
{\tt hep-th/0306087}
 }
\end{flushright}

\vspace*{0.1cm}

\begin{center}
 \Large\bf Partition Function and Open/Closed String Duality \\
in Type IIA String Theory on a PP-wave
\end{center}
\vspace*{0.5cm}
\centerline{\large Hyeonjoon Shin$^{1,3a}$,
Katsuyuki Sugiyama$^{2b}$
and Kentaroh Yoshida$^{3c}$}

\begin{center}
$^{1}$\emph{BK21 Physics Research Division and Institute of Basic
 Science,\\
Sungkyunkwan University, Suwon 440-746, Korea.} \\
\vspace*{0.2cm}
$^{2}$\emph{Department of Physics, Kyoto University, 
Kyoto 606-8501, Japan.} \\
\vspace{0.2cm}
$^{3}$\emph{Theory Division, High Energy Accelerator Research 
Organization (KEK),\\
Tsukuba, Ibaraki 305-0801, Japan.} 
\\
\vspace*{1cm}
\parbox{13cm}{{\tt E-mail:~$^{a}$hshin@newton.skku.ac.kr 
(hshin@post.kek.jp) \\
\hspace{1.5cm} $^{b}$sugiyama@phys.h.kyoto-u.ac.jp\\
\hspace{1.5cm} $^{c}$kyoshida@post.kek.jp}}
\end{center}

\vfill

\centerline{\bf Abstract}

We discuss partition functions of $\mathcal{N}=(4,4)$ type IIA string
theory on the pp-wave background. This theory is shown to be modular
invariant.  The boundary states are constructed and possible D-brane
instantons are classified.  Then we calculate cylinder amplitudes in
both closed and open string descriptions and check the open/closed
string duality.  Furthermore we consider general properties of modular
invariant partition functions in the case of pp-waves.

\vspace*{1.5cm}

\thispagestyle{empty}
\setcounter{page}{0}

\newpage 

\section{Introduction}

Recently, superstring theories on pp-waves have been very focused upon.
The maximally supersymmetric pp-wave type solution in eleven dimensions
\cite{KG} has been known for a long time while the maximally
supersymmetric type IIB pp-wave solution was found in \cite{BFHP1} in
the recent progress.  It was also pointed out in \cite{BFHP2} that this
pp-wave background is related to the $AdS$-geometry via the Penrose
limit \cite{P,G}. Then, the type IIB superstring theory on the maximally
supersymmetric pp-wave background was constructed \cite{M,MT}.  By using
this pp-wave superstring theory, the study of $AdS$/CFT correspondence
has greatly proceeded \cite{BMN}. In particular, the $AdS$/CFT
correspondence has been studied at the stringy level beyond the
supergravity analysis.
 
In the study of pp-wave backgrounds, the matrix model on the pp-wave,
which was proposed by Berenstein-Maldacena-Nastase \cite{BMN}, has been
much studied.  This matrix model is closely related to a supermembrane
theory on the pp-wave background \cite{DSR,SY1}.  We have discussed the
supermembrane theory and matrix model on the pp-wave from the several
aspects \cite{SY1,HS1,SY2,SY3,NSY}.  In particular, we showed the
correspondence of brane charges in the supermembrane theory \cite{SY1}
and matrix model \cite{HS1} in the pp-wave case as well as in flat space
\cite{BSS}.

A supermembrane in eleven dimensions is related to a string in ten
dimensions via the double dimensional reduction. We constructed the type
IIA pp-wave background with 24 supersymmetries and string theory on this
pp-wave background \cite{SY4,HS2}, which is called $\mathcal{N}=(4,4)$
type IIA string theory on the pp-wave.  We discussed the classification
of the allowed D-branes \cite{SY4,HS3} by following the work of
Dabholkar and Parvizi \cite{DP}.  After these works, the covariant
classification of D-branes was done in \cite{HPS} where D0-branes could
be studied.  In addition, the spectrum of this type IIA string theory
was compared to fluctuations of the linearized type IIA supergravity
around the pp-wave background\cite{KS}.  The thermodynamics of this type
IIA string theory was recently studied in \cite{HPY} \footnote{Section 2
has some overlap with the work \cite{HPY}.  Thermal partition function
is discussed in the cases of other pp-wave strings \cite{S1,S2}. }.

On the other hand, as an important and interesting subject, the modular
invariance of string theories on pp-waves has been studied by several
authors \cite{BGG,GG,T,S1,S2,HPY,GGSS}.  It has been turned out that
these theories are modular invariant in spite of mass terms in the
action. In this paper, motivated by the previous works, we will be
interested in the $\mathcal{N}=(4,4)$ type IIA string theory on the
pp-wave background obtained in \cite{SY4,HS2} and study its partition
function and open/closed string duality.  The pp-wave background we will
consider is not maximally supersymmetric but has 24
supersymmetries. Hence it is interesting to study whether the modular
invariance and consistency condition between open and closed strings are
satisfied or not in such less supersymmetric case. Moreover, since the
number of preserved supersymmetries is nontrivial even in the case of
supersymmetric D-branes, it is also interesting to construct 
boundary states in our model\footnote{We note that, contrary to the
present type IIA case, the boundary states in the type IIB string theory
on the pp-wave background have been relatively much
studied\cite{BP,BGG,Green,GGSS}.}.

In this paper we discuss partition functions in the $\mathcal{N}=(4,4)$
type IIA string theory on the pp-wave.  We first show that our theory is
modular invariant. The boundary states is constructed and then we
classify the allowed D-brane instantons. They are 1/2 BPS states
(preserving 12 supersymmetries) at the origin and 1/3 BPS ones
(preserving 8 supersymmetries) away from the origin.  This result is
consistent with the classification of D-branes.  Then we calculate the
cylinder amplitude in the closed and open string descriptions and study
the open/closed string channel duality in our theory. Finally, we
discuss general properties of modular invariant partition function and
classify several models.

This paper is organized as follows: In section 2 we prove the modular
invariance of $\mathcal{N}=(4,4)$ type IIA string theory on the pp-wave
background.  The Witten index of this theory is shown to be one.
Section 3 is devoted to a brief review about the supersymmetries of our
theory. In section 4 we construct boundary states and classify D-brane
instantons.  In section 5 we calculate the cylinder amplitude in the
closed string description. In section 6 the amplitude is derived in
terms of open string and the open/closed string channel duality in our
theory is proven. From the channel duality, the normalization factor of
the boundary state is determined.  In section 7, based on the result of
section 2, we discuss several properties of modular invariant partition
function in some general setup. Finally, section 8 is devoted to 
conclusions and discussions.

\section{Modular Invariance of Type IIA String Theory}

In this section we will discuss the modular invariance of type IIA 
string theory in the closed string description.   

The action of our type IIA string in the light-cone gauge is given by 
\begin{eqnarray}
 S_{\rm closed} &=& \frac{1}{4\pi\al'}
\int\! d\tau\!\! \int^{2\pi}_0\!\!\!d\sigma\, \Biggl[
\sum_{i=1}^8\partial_+ x^i\partial_- x^i - 
\left(\frac{\mu}{3}\right)^2 \sum_{a=1}^4 (x^a)^2 
- \left(\frac{\mu}{6}\right)^2\sum_{b=5}^8(x^b)^2 \Biggr]  \\ 
&& + 
\frac{i}{2\pi}
\int\! d\tau\!\! \int^{2\pi}_0\!\!\!d\sigma\, \Biggl[
\Psi^{1+}{}^{\s T}\partial_-\Psi^{1+} 
+ \Psi^{1-}{}^{\s T}\partial_-\Psi^{1-} 
+ \Psi^{2+}{}^{\s T}\partial_+\Psi^{2+} 
+ \Psi^{2-}{}^{\s T}\partial_+\Psi^{2-} \nn \\
&& 
- \frac{\mu}{3}\Psi^{1-}{}^{\s T}\Pi^{\s T}\Psi^{2+} 
+ \frac{\mu}{3}\Psi^{2+}{}^{\s T}\Pi\Psi^{1-}
- \frac{\mu}{6}\Psi^{1+}{}^{\s T}\Pi^{\s T}\Psi^{2-} 
+ \frac{\mu}{6}\Psi^{2-}{}^{\s T}\Pi\Psi^{1+} \Biggr]
\,, \nn
\end{eqnarray}
where $\al'$ is a string tension and we have set $p^+=1$. The
$\gamma^r$'s are 16 $\times$ 16 $SO(9)$ gamma matrices and we defined
$\Pi \equiv \gamma^{123}$ $(\Pi^{\s T}\equiv \gamma^{321})$.  Each of
the spinors $\Psi^{i\pm}~~(i=1,2)$ has four independent components and 
the superscript $\pm$ represents the chirality measured by
$\gamma^{1234}$: $ \gamma^{1234}\Psi^{i\pm} = \pm 1\cdot \Psi^{i\pm}\,.$
This theory has 24 supersymmetries (8 dynamical supersymmetries and 16
kinematical supersymmetries). 
The equations of motion are described by 
\begin{eqnarray}
\label{em1}
&& \partial_+ \partial_- x^a + \frac{\mu^2}{9}\,x^a = 0 \quad 
(a=1,2,3,4)\,,  \\
\label{em2}
&& \partial_+ \partial_- x^b + \frac{\mu^2}{36}\,x^b = 0 \quad
(b=5,6,7,8)\,,  \\
\label{em3}
&&  \partial_+ \Psi^{2+} + \frac{\mu}{3}\,\Pi\,\Psi^{1-} 
= 0\,, \quad \partial_- \Psi^{1-} - \frac{\mu}{3}\,\Pi^{\s T}\,
\Psi^{2+} = 0\,,  \\ 
&&  \partial_+ \Psi^{2-} + \frac{\mu}{6}\,\Pi\,\Psi^{1+} 
= 0\,, \quad 
\partial_- \Psi^{1+} - \frac{\mu}{6}\,\Pi^{\s T}\,
\Psi^{2-} = 0\,. \label{em4} 
\end{eqnarray}
By solving the equations of motion (\ref{em1}) and (\ref{em2}), 
we can obtain 
the mode-expansions of bosonic variables represented by 
\begin{eqnarray}
x^a(\tau,\,\sigma) &=&  
x_0^{a} \cos\left(\frac{\mu}{3}\tau\right) 
+ \left(
\frac{3}{\mu}\right)\al' p_0^{a}\sin\left(\frac{\mu}{3}\tau\right) 
+ i\sqrt{\frac{\al'}{2}}\sum_{n\neq 0}\frac{1}{\omega_n}
\left[\alpha_n^{a}\phi_n + \bar{\alpha}_n^{a} 
\tilde{\phi}_n \right]\,,  \\
x^b(\tau,\,\sigma) &=& 
x_0^{b} \cos\left(\frac{\mu}{6}\tau\right) 
+ \left(\frac{6}{\mu}\right)
\al' p_0^{b}\sin\left(\frac{\mu}{6}\tau\right) 
+ i\sqrt{\frac{\al'}{2}}\sum_{n\neq 0}\frac{1}{\omega_n'}
\left[
\alpha_n^{b}\phi_n' + \bar{\alpha}_n^b\tilde{\phi}_n'  
\right]\,. 
\end{eqnarray}
From the equations of motion (\ref{em3}) 
and (\ref{em4}),  
the mode-expansions of fermionic variables are represented by
\begin{eqnarray}
 \Psi^{1-}(\tau,\,\sigma) &=& 
 \Pi^T\Psi_0\sin\left(\frac{\mu}{3}\tau\right) 
-\Pi^T\widetilde{\Psi}_0\cos\left(\frac{\mu}{3}\tau\right) \nn \\
&& \qquad + \sum_{n\neq 0}c_n\left(\frac{3}{\mu}i(\omega_n - n)
\,\Pi^{\s T}
\Psi_n\phi_n + \widetilde{\Psi}_n\tilde{\phi}_n\right) 
\,, \\
\Psi^{2+}(\tau,\,\sigma) &=& 
\Psi_0 \cos\left(\frac{\mu}{3}\tau\right) 
+ \widetilde{\Psi}_0\sin\left(\frac{\mu}{3}\tau\right) \nn \\
&& \qquad + \sum_{n\neq 0}
c_n\left[\Psi_n\phi_n - \frac{3}{\mu}i(\omega_n -n)
\,\Pi\widetilde{
\Psi}_n\tilde{\phi}_n \right] \,,  \\
 \Psi^{1+}(\tau,\,\sigma) &=& 
\Pi^T\Psi_0'\sin\left(\frac{\mu}{6}\tau\right) 
- \Pi\widetilde{\Psi}_0'\cos\left(\frac{\mu}{6}\tau\right) \nn \\
&& \qquad + \sum_{n\neq 0}c_n'\left(\frac{6}{\mu}i(\omega_n' - n)
\,\Pi^{\s T}
\Psi_n'\phi_n' + \widetilde{\Psi}_n'\tilde{\phi}_n'
\right) \,, \\
\Psi^{2-}(\tau,\,\sigma) &=& 
\Psi_0' \cos\left(\frac{\mu}{6}\tau\right) 
+ \widetilde{\Psi}_0'\sin\left(\frac{\mu}{6}\tau\right) \nn \\
&& \qquad + \sum_{n\neq 0}
c_n'\left[\Psi_n'\phi_n' - \frac{6}{\mu}i(\omega_n' -n)
\,\Pi\widetilde{\Psi}_n'\tilde{\phi}_n'\right] \,.  
\end{eqnarray} 
Here we have introduced several notations:
\begin{eqnarray}
&& \omega_n 
\;=\; {\rm sgn}(n)\sqrt{n^2 + \left(\frac{\mu}{3}\right)^2}\,,\quad 
\omega_n' \;=\; 
{\rm sgn}(n)\sqrt{n^2 + \left(\frac{\mu}{6}\right)^2 }
\,, \label{omega} \nn\\ 
&& \phi_n 
\;=\; \exp\left(-i(\omega_n\tau - n\sigma)\right)\,,\quad 
\tilde{\phi}_n 
\;=\; \exp\left(-i\left(\omega_n \tau  + n\sigma\right)
\right)\,, \nn \\
&& \phi_n' 
\;=\; \exp\left(-i(\omega_n'\tau - n\sigma)\right)\,,\quad 
\tilde{\phi}_n' 
\;=\; \exp\left(-i\left(\omega_n'\tau  + n\sigma\right)
\right)\,, \nn \\
&& c_n \;=\; \left(1 + \left(\frac{3}{\mu}\right)^2(
\omega_n - n)^2\right)^{-1/2}\,, \quad 
c_n' \;=\; \left(1 + \left(\frac{6}{\mu}\right)^2(
\omega_n' - n)^2\right)^{-1/2}
\,. \nn 
\end{eqnarray}
Now we shall quantize the theory by imposing (anti)commutation relations.  
The commutation relations for bosonic modes are given by 
\begin{eqnarray} 
\label{com-b}
\left[x_0^i,\,p^j_0\right] 
&=& i\del^{ij}\,, \quad 
\left[\bar{\al}^i_m, \, \al_n^j\right] 
\;=\; [\al^i_m,\,\bar{\al}^j_n] \;=\; 0  \quad (i,j = 1,\ldots, 8)\,,\nn 
\\  
\left[\al_m^a,\, \alpha_n^{a'}\right] &=& 
\left[\bar{\al}_m^a,\,\bar{\al}_n^{a'}\right] 
\;=\; \omega_m\del_{m+n,0}\,\del^{aa'} 
\quad (a,\,a' = 1,2,3,4)\,, \nn \\ 
\left[\al_m^b,\, \alpha_n^{b'}\right] &=& 
\left[\bar{\al}_m^b,\,\bar{\al}_n^{b'}\right] \;=\;
\omega_m'\del_{m+n,0}\,\del^{bb'}  
\quad (b,\,b' = 5,6,7,8)\,, \nn 
\end{eqnarray}
and the anticommutation relations for fermionic modes are written as 
\begin{eqnarray}
\label{com-f}
\{(\Psi_m)_{\al},\, (\widetilde{\Psi}_n)_{\beta}^{\s T}
\} &=& 
\{
(\widetilde{\Psi}_m)_{\al},\,(\Psi_n)_{\beta}^{\s T}
\} \; = \; 0\,,  \nn \\
\{(\Psi_m)_{\al},\,(\Psi_n)_{\beta}^{\s T}\}
&=& 
\{(\widetilde{\Psi}_m)_{\al},\,(\widetilde{\Psi}_n)_{\beta}^{\s T} 
\} \;=\;  \frac{1}{2}\del_{m+n,0}\,\del_{\al\beta}\,, \nn \\
\{(\Psi_m')_{\al},\, (\widetilde{\Psi}_n')_{\beta}^{\s T}
\} &=& 
\{
(\widetilde{\Psi}_m')_{\al},\,(\Psi_n')_{\beta}^{\s T}
\} \; = \; 0\,, \nn \\
\{(\Psi_m')_{\al},\,(\Psi_n')_{\beta}^{\s T}\}
&=& \{(\widetilde{\Psi}_m')_{\al},\,(\widetilde{\Psi}_n')_{\beta}^{\s T} 
\} \;=\;  \frac{1}{2}\del_{m+n,0}\,\del_{\al\beta}\,. \nn 
\end{eqnarray}
Now let us introduce the annihilation and creation operators:
 \begin{eqnarray}
&&a^a_0 \equiv  \sqrt{\frac{\alpha'}{2}}
\sqrt{\frac{3}{\mu}}
\left(p^a_0 - i\frac{\mu}{3\alpha'}x^a_0\right)\,,\quad 
{a}^{a\dagger}_0 \equiv \sqrt{\frac{\alpha'}{2}}
\sqrt{\frac{3}{\mu}}
\left(p^a_0 + i\frac{\mu}{3\alpha'}x^a_0\right)\,\quad 
(a=1,2,3,4)\,, \nn \\
&&a^a_n \equiv \frac{1}{\sqrt{\omega_n}}\alpha^a_n\,,\quad 
a^{a\dagger}_n \equiv \frac{1}{\sqrt{\omega_n}}\alpha^a_{-n}\,,\quad  
\bar{a}^a_n \equiv \frac{1}{\sqrt{\omega_n}}\bar{\alpha}^a_n\,,\quad 
\bar{a}^{a\dagger}_n \equiv 
\frac{1}{\sqrt{\omega_n}}\bar{\alpha}^a_{-n} \quad 
(n>0)\,, \nn 
\end{eqnarray}
for the sector with mass $\mu/3$ and 
\begin{eqnarray}
&& a^b_0 \equiv \sqrt{\frac{\alpha'}{2}}
\sqrt{\frac{6}{\mu}}
\left(p^a_0 - i\frac{\mu}{6\alpha'}x^a_0\right)\,,\quad 
{a}^{b\dagger}_0 \equiv \sqrt{\frac{\alpha'}{2}}
\sqrt{\frac{6}{\mu}}
\left(p^a_0 + i\frac{\mu}{6\alpha'}x^a_0\right) \quad (b=5,6,7,8)\,,
\nn \\
&&a^b_n=\frac{1}{\sqrt{\omega'_n}}\alpha^b_n\,, \quad 
a^{b\dagger}_n=\frac{1}{\sqrt{\omega'_n}}\alpha^b_{-n}\,,\quad 
\bar{a}^b_n=\frac{1}{\sqrt{\omega'_n}}\bar{\alpha}^b_n\,, \quad 
\bar{a}^{b\dagger}_n=\frac{1}{\sqrt{\omega'_n}}\bar{\alpha}^b_{-n} \quad 
(n>0)\,, \nn
\end{eqnarray}
for the sector with mass $\mu/6$, and those for fermionic variables:  
\begin{eqnarray}
&& S_0=\Psi_0+i\tilde{\Psi}_0\,,\quad 
S^{\dagger}_0=\Psi_0-i\tilde{\Psi}_0\,, \nn\\
&& S_n=\sqrt{2}\Psi_n\,,\quad 
S^{\dagger}_n=\sqrt{2}\Psi_{-n}\,,\quad 
\tilde{S}_n=\sqrt{2}\tilde{\Psi}_n\,,\quad 
\tilde{S}^{\dagger}_n=\sqrt{2}\tilde{\Psi}_{-n}\,,\qquad (n>0) \nn \\
&& S'_n=\sqrt{2}\Psi'_n\,,\quad 
{S'}^{\dagger}_n=\sqrt{2}\Psi'_{-n}\,,\quad 
\tilde{S}'_n=\sqrt{2}\tilde{\Psi}'_n\,,\quad 
\tilde{S'}^{\dagger}_n=\sqrt{2}\tilde{\Psi'}_{-n}\,.\qquad (n>0)\nn
\end{eqnarray}
Then the commutation relations are rewritten as 
\begin{eqnarray}
&&[a^a_m,a^{a'\dagger}_n]=\delta^{aa'}\delta_{m,n}\,, \quad 
[\bar{a}^a_m,\bar{a}^{a'\dagger}_n]=\delta^{aa'}\delta_{m,n}\,
\qquad (a,\,a'=1,2,3,4)\,, \nn\\
&&[a^b_m,a^{b'\dagger}_n]=\delta^{bb'}\delta_{m,n}\,, \quad 
[\bar{a}^b_m,\bar{a}^{b'\dagger}_n]=\delta^{bb'}\delta_{m,n}\,
\qquad (b,\,b'=5,6,7,8)\,, \nn
\end{eqnarray}
and the anticommutation relations are given by 
\begin{eqnarray}
&&\{(S_m)_{\alpha},(S^{\dagger}_n)_{\beta}\}=
\delta_{\alpha\beta}\delta_{m,n}\,,\quad 
\{(\tilde{S}_m)_{\alpha},(\tilde{S}^{\dagger}_n)_{\beta}\}
=\delta_{\alpha\beta}\delta_{m,n}\,,\nn\\
&&\{({S'}_m)_{\alpha},({S'}^{\dagger}_n)_{\beta}\}=
\delta_{\alpha\beta}\delta_{m,n}\,, \quad 
\{(\tilde{S}'_m)_{\alpha},(\tilde{S'}^{\dagger}_n)_{\beta}\}
=\delta_{\alpha\beta}\delta_{m,n}\,.\nn
\end{eqnarray}
By using the above (anti)commutation relations, we can represent the 
Hamiltonian and momentum 
in terms of creation and annihilation operators as follows: 
\begin{eqnarray}
H &=& 
\sum^{\infty}_{n=-\infty}\left(\omega_n N_n +\omega'_n N'_n\right)\,, \quad 
  P=\sum^{\infty}_{n=-\infty}(nN_n+nN'_n)\,. 
\end{eqnarray}
where $N_n$ and $N_n'$ are defined by 
\begin{eqnarray}
&& N_0 \equiv \sum^4_{a=1}a^{a\dagger }_0a^a_0+S^{\dagger}_0S_0\,,\quad 
N'_0 \equiv \sum^8_{b=5}a^{b\dagger }_0a^b_0+{S'}^{\dagger}_0S'_0\,,
\nn\\
&& N_n \equiv \sum^4_{a=1}a^{a\dagger}_na^a_n+S^{\dagger}_nS_n\,,\quad 
N'_n \equiv \sum^8_{b=5}a^{b\dagger}_na^b_n+{S'}^{\dagger}_nS'_n 
\qquad (n>0)\,,\nn \\
&& N_{-n} \equiv \sum^4_{a=1}\bar{a}^{a\dagger}_n\bar{a}^a_n
+\tilde{S}^{\dagger}_n\tilde{S}_n\,,\quad 
N'_{-n} \equiv \sum^8_{b=5}\bar{a}^{b\dagger}_n\bar{a}^b_n
+\tilde{S'}^{\dagger}_n\tilde{S}'_n
\qquad (n>0)\,. \nn
\end{eqnarray} 
Now we shall introduce the Casimir Energy defined by 
\begin{eqnarray}
&& \Delta (\nu;a) \;\equiv\; \frac{1}{2}\sum_{n\in {\bf Z}}
\sqrt{\nu^2+(n+a)^2}-\frac{1}{2}\int^{+\infty}_{-\infty}\!dk
\,\sqrt{\nu^2+k^2}\nn\\
&=& -\frac{1}{2\pi^2}\sum_{\ell \geq 1}\cos (2\pi a\ell)
\int^{\infty}_0 \!\!\! ds\, e^{-\ell^2 s-\frac{\pi^2\nu^2}{s}}  
= -\frac{1}{2}\int^{\infty}_0 \!\!\!ds\, e^{-\frac{\nu^2}{s}}
\left(\theta_3(i\pi s,a)-1\right)\,. 
\end{eqnarray}
Then we can express the vacuum energy for eight bosons as 
\begin{eqnarray}
E^0_B &=&\sum^4_{a=1}\left(\frac{1}{2}\sum_{n\in {\bf Z}}\omega_n\right)
+\sum^8_{b=5}\left(\frac{1}{2}\sum_{n\in {\bf Z}}\omega'_n\right) 
\;\cong\; 4(\Delta (\mu/3 ;0)+
\Delta (\mu/6 ;0) )\,,
\end{eqnarray}
and that for fermions as 
\begin{eqnarray}
E^0_F &=& - 4\left(
\frac{1}{2}\sum_{n\in {\bf Z}}\omega_n+
\frac{1}{2}\sum_{n\in {\bf Z}}\omega'_n
\right) \;\cong\; - 4
\left[\Delta (\mu/3 ;0)+\Delta (\mu/6 ;0)\right]\,,
\end{eqnarray}
where the symbol $\cong$ means the equality after the regularization of
zero-point energies and the factor 4 appears since each fermion
considered here has four independent components.

Here let us evaluate the toroidal partition function:  
\begin{eqnarray}
&&Z= \mbox{Tr}\left[
(-1)^{\bf F}e^{-2\pi \tau_2 H+2\pi i\tau_1 P}
\right]\,,
\end{eqnarray}
where $\tau_1$ and $\tau_2$ are modular parameters and 
${\bf F}$ is the fermion number operator. 

First we will consider a partition function for one boson with mass $\nu$. 
The number operator, Hamiltonian and momentum are given by 
\begin{eqnarray}
&&N_0=a^{\dagger}_0a_0\,,\,\,\,
N_n=a^{\dagger}_na_n\,,\,\,\,
N_{-n}=\bar{a}^{\dagger}_n\bar{a}_n \quad (n>0)\,, \nn \\
&& H = \omega_0 N_0 + \sum_{n=1}^{\infty}\left(
\omega_nN_n + \omega_nN_{-n}\right)\,,\quad 
P=\sum_{n=1}^{\infty}\left( nN_n - nN_{-n}\right)\,, \nn 
\end{eqnarray}
and so we can obtain the partition function: 
\begin{eqnarray}
&&Z=\left[\Theta_{(0,0)}(\tau ,\bar{\tau};\nu)\right]^{-1/2}\,, 
\end{eqnarray}
where we have introduced the `massive' theta function defined by 
\begin{eqnarray}
\Theta_{(a,b)}(\tau ,\bar{\tau};\nu)
\equiv e^{4\pi\tau_2 \Delta (\nu ;a)}
\prod_{n\in {\bf Z}}\left|
1-e^{-2\pi \tau_2 \sqrt{\nu^2 +(n+a)^2}+2\pi \tau_1 (n+a)+2\pi ib}
\right|^2\,.
\label{theta}
\end{eqnarray}
Next we consider a partition function for 
single component 
fermion. The number operator, Hamiltonian and momentum are given by 
\begin{eqnarray}
&&N_0=S^{\dagger}_0S_0\,,\quad 
N_n=S^{\dagger}_nS_n\,,\quad 
N_{-n} = \tilde{S}^{\dagger}_n\tilde{S}_n \quad (n>0)\,, \nn \\
&& H = \omega_0N_0 + \sum_{n>0}\left(
\omega_nN_n + \omega_nN_{-n}\right)\,,\quad 
P = \sum_{n>0}\left( nN_n - nN_{-n}\right)\,,\nn  
\end{eqnarray}
and hence we can obtain the partition function: 
\begin{eqnarray}
&&Z=\left[\Theta_{(0,0)}(\tau ,\bar{\tau};\nu)\right]^{+1/2}\,. 
\end{eqnarray}
If we recall the field contents of our model: 
\begin{eqnarray}
&&\mbox{bosons}\left\{
\begin{array}{l}
x^a ~~(a=1,2,3,4)  \\
x^b~~(b=5,6,7,8) 
\end{array}
\right.\,,\quad 
\mbox{fermions}\left\{
\begin{array}{llll}
(S,\tilde{S}) & & (\mbox{mass}~~\frac{\mu}{3} \mbox{~~sector})\\
(S',\tilde{S}') & & (\mbox{mass}~~\frac{\mu}{6} \mbox{~~sector}) \\
\end{array}
\right.\,,
\end{eqnarray}
the bosonic and fermionic partition functions $Z_B$ and $Z_F$ 
are given by 
\begin{eqnarray}
Z_B &=& \left[\Theta_{(0,0)}(\tau ,\bar{\tau};\mu/3)\right]^{-2}
\left[\Theta_{(0,0)}(\tau ,\bar{\tau};\mu/6)\right]^{-2}\,, \nn \\
Z_F&=&\left[\Theta_{(0,0)}(\tau ,\bar{\tau};\mu/3)\right]^{+2}
\left[\Theta_{(0,0)}(\tau ,\bar{\tau};\mu/6)\right]^{+2}\,, \nn
\end{eqnarray}
and hence the total partition function is given by 
\begin{eqnarray}
&& Z=Z_B \cdot Z_F=1 \,. 
\label{total}
\end{eqnarray}
Thus we have shown that our theory is modular invariant at the one-loop
level since the total partition function $Z$ of (\ref{total}) is
independent of the modular parameters $\tau$ and $\bar{\tau}$.  As will
be shown in more detail with some generality in the section 7, this
result implies that the Witten index is one as in the cases of other
string theories on pp-waves.  It should be remarked that our type IIA
string theory is modular invariant in the sector with mass $\mu/3$ and
in that with $\mu/6$, respectively.

\section{Supersymmetries of Type IIA String Theory}

In this section we will briefly review about $\mathcal{N}=(4,4)$
supersymmetries of the type IIA string theory on the pp-wave background
\cite{HS2,HS3}, according to which the world-sheet variables are
arranged into two supermultiplets, ($x^a,\Psi^{1-},\Psi^{2+}$) and
($x^b,\Psi^{1+},\Psi^{2-}$).  Then we will rewrite the supercharges in
terms of modes in order to construct the boundary states in
section 4.
   
This type IIA string theory on the pp-wave background has 24
supersymmetries, among which 8 are dynamical and 16 are kinematical.  
The dynamical supersymmetry transformation laws for the multiplet
($x^a,\Psi^{1-},\Psi^{2+}$) with mass $\mu/3$ are given by
\begin{eqnarray}
\del x^a &=& 2i\al'\left(\Psi^{1-T}\gamma^a\epsilon^{1+} 
+ \Psi^{2+T}\gamma^a\epsilon^{2-}\right)\,,  \\
\del \Psi^{1-} &=& \partial_+ x^a \gamma^a\epsilon^{1+} 
+ \frac{\mu}{3}x^a\gamma^4\gamma^a\epsilon^{2-}\,, \quad 
\del \Psi^{2+} \;=\; \partial_- x^a \gamma^a\epsilon^{2-} 
- \frac{\mu}{3}x^a\gamma^4\gamma^a\epsilon^{1+}\,, \nn 
\end{eqnarray}
and those for the multiplet ($x^b,\Psi^{1+},\Psi^{2-}$) with mass $\mu/6$, 
are written as 
\begin{eqnarray}
\del x^b &=& 2i\al'\left(\Psi^{1+T}\gamma^b\epsilon^{2-} 
+ \Psi^{2-T}\gamma^b\epsilon^{1+}\right)\,,  \\
\del \Psi^{1+} &=& \partial_- x^b \gamma^b \epsilon^{2-} 
+ \frac{\mu}{6}x^b\gamma^4\gamma^b\epsilon^{1+}\,, \quad 
\del \Psi^{2-} \;=\; \partial_+ x^b \gamma^b \epsilon^{1+} 
- \frac{\mu}{6} x^b\gamma^4\gamma^b\epsilon^{2-}\,, \nn
\end{eqnarray} 
where the constant spinors $\epsilon^{1+}$ and $\epsilon^{2-}$ 
satisfy the following chirality conditions:  
\begin{eqnarray}
&& \gamma^9\epsilon^{1+} = + \epsilon^{1+}\,, \quad 
\gamma^{1234}\epsilon^{1+} = +\epsilon^{1+}\,, \quad 
\gamma^9\epsilon^{2-} = - \epsilon^{2-}\,, \quad 
\gamma^{1234}\epsilon^{2-} = - \epsilon^{2-}\,,
\end{eqnarray} 
respectively. As for the kinematical supersymmetry, the 
transformation laws are given by 
$\tilde{\del}x^a = \tilde{\delta}x^b =0$ and 
\begin{eqnarray}
\tilde{\delta}\Psi^{1-} 
&=& \cos\left(\frac{\mu}{3}\tau\right)\tilde{\epsilon}^{1-} 
- \sin\left(\frac{\mu}{3}\tau\right)\gamma^{123}\tilde{\epsilon}^{2+}\,, \\
\tilde{\delta}\Psi^{2+}
&=& \cos\left(\frac{\mu}{3}\tau\right)\tilde{\epsilon}^{2+} 
- \sin\left(\frac{\mu}{3}\tau\right)\gamma^{123}\tilde{\epsilon}^{1-}\,,\nn \\
\tilde{\delta}\Psi^{1+} 
&=& \cos\left(\frac{\mu}{6}\tau\right)\tilde{\epsilon}^{1+} 
- \sin\left(\frac{\mu}{6}\tau\right)\gamma^{123}\tilde{\epsilon}^{2-}\,, \\
\tilde{\delta}\Psi^{2-}
&=& \cos\left(\frac{\mu}{6}\tau\right)\tilde{\epsilon}^{2-} 
- \sin\left(\frac{\mu}{6}\tau\right)\gamma^{123}\tilde{\epsilon}^{1+}\,,\nn
\end{eqnarray}
where the constant spinors $\tilde{\epsilon}^{1+}$, 
$\tilde{\epsilon}^{1-}$, $\tilde{\epsilon}^{2+}$ and 
$\tilde{\epsilon}^{2-}$ satisfy the chirality conditions  
in terms of $\gamma^9$ and $\gamma^{1234}$ in the same way as 
the dynamical supersymmetry case.  

By the use of Noether's theorem, we can construct the associated
supercharges. Firstly the dynamical supercharges are obtained as
\begin{eqnarray}
\epsilon^{\s T}Q_{(\mu/3)} = 
\epsilon^{1+{\s T}}Q^{1+} + \epsilon^{2-{\s T}}Q^{2-}\,, \quad 
\epsilon'{}^{\s T}Q_{(\mu/6)} = 
\epsilon^{2-{\s T}}Q'{}^{2-} + \epsilon^{1+{\s T}}Q'{}^{1+}\,.
\end{eqnarray}
Here the $Q^{1+}$ and $Q^{2-}$ are the quantities defined as 
\begin{eqnarray}
Q^{1+} &\equiv& - \frac{i}{2\pi}\int^{2\pi}_0\!\!\!\!d\sigma\,
\left[
\partial_+ x^a \gamma^a\Psi^{1-} - \frac{\mu}{3}x^a\gamma^a\gamma^4\Psi^{2+}
\right]\,, \\
Q^{2-} &\equiv& - \frac{i}{2\pi}\int^{2\pi}_0\!\!\!\!d\sigma\,
\left[
\partial_- x^a \gamma^a\Psi^{2+} + \frac{\mu}{3}x^a\gamma^a\gamma^4\Psi^{1-}
\right]\,,  
\end{eqnarray}
and, for the $Q'{}^{2-}$ and $Q'{}^{1+}$,  
\begin{eqnarray}
Q'{}^{2-} &\equiv& - \frac{i}{2\pi}\int^{2\pi}_0\!\!\!\!d\sigma\,
\left[
\partial_+ x^b \gamma^b\Psi^{1+} - \frac{\mu}{6}x^b\gamma^b\gamma^4\Psi^{2-}
\right]\,, \\
Q'{}^{1+} &\equiv& - \frac{i}{2\pi}\int^{2\pi}_0\!\!\!\!d\sigma\,
\left[
\partial_- x^b \gamma^b\Psi^{2-} + \frac{\mu}{6}x^b\gamma^b\gamma^4\Psi^{1+}
\right]\,. 
\end{eqnarray}
Secondly, the kinematical supercharges are obtained as 
\begin{eqnarray}
\tilde{Q}_{(\mu/3)} \equiv \tilde{\epsilon}^{1-{\s T}}\tilde{Q}^{1-} 
+ \tilde{\epsilon}^{2+{\s T}}\tilde{Q}^{2+}\,, \quad 
\tilde{Q}_{(\mu/6)} \equiv \tilde{\epsilon}^{1+{\s T}}\tilde{Q}^{1+} 
+ \tilde{\epsilon}^{2-{\s T}}\tilde{Q}^{2-}\,, 
\end{eqnarray}
where the $\tilde{Q}^{1-}$ and $\tilde{Q}^{2+}$ are defined by 
\begin{eqnarray}
\tilde{Q}^{1-} &\equiv& \frac{i}{2\pi}\int^{2\pi}_0\!\! d\sigma\,
\left[
\cos\left(\frac{\mu}{3}\tau\right)\Psi^{1-} 
+ \sin\left(\frac{\mu}{3}\tau\right)\gamma^{123}\Psi^{2+}
\right]\,, \\
\tilde{Q}^{2+} &\equiv& \frac{i}{2\pi}\int^{2\pi}_0\!\! d\sigma\,
\left[
\cos\left(\frac{\mu}{3}\tau\right)\Psi^{2+} 
+ \sin\left(\frac{\mu}{3}\tau\right)\gamma^{123}\Psi^{1-}
\right]\,,
\end{eqnarray}
and, for the $\tilde{Q}^{1-}$ and $\tilde{Q}^{2+}$, 
\begin{eqnarray}
\tilde{Q}^{1+} &\equiv& \frac{i}{2\pi}\int^{2\pi}_0\!\! d\sigma\,
\left[
\cos\left(\frac{\mu}{6}\tau\right)\Psi^{1+} 
+ \sin\left(\frac{\mu}{6}\tau\right)\gamma^{123}\Psi^{2-}
\right]\,, \\
\tilde{Q}^{2-} &\equiv& \frac{i}{2\pi}\int^{2\pi}_0\!\! d\sigma\,
\left[
\cos\left(\frac{\mu}{6}\tau\right)\Psi^{2-} 
+ \sin\left(\frac{\mu}{6}\tau\right)\gamma^{123}\Psi^{1+}
\right]\,.
\end{eqnarray}

Now we can rewrite the above supercharges in terms of 
creation and annihilation operators by inserting the mode-expansions 
of bosonic and fermionic degrees of freedom. 
\begin{eqnarray}
\sqrt{\al'}^{-1}Q^{1+} &=& 
c_0 \sqrt{\frac{\mu}{3}} \left(a_0^{a\dagger}\gamma^a\Pi^{\s T}S_0 
- a_0^a\gamma^a\Pi^{\s T}S_0^{\dagger}\right) 
-i \sum_{n=1}^{\infty}c_n\sqrt{\omega_n}\left(
\bar{a}_n^{a\dagger}\gamma^a\tilde{S}_n + \bar{a}_n^a \gamma^a\tilde{S}_n^{\dagger}
\right)
\nn \\
&& \quad + \frac{1}{2}\cdot\frac{\mu}{3}\sum_{n=1}^{\infty}
\frac{1}{c_n\sqrt{\omega_n}}\left(
a_n^{a\dagger}\gamma^a\Pi^{\s T}S_n 
- a_n^a \gamma^a\Pi^{\s T}S_n^{\dagger}
\right)\,, \\
\sqrt{\al'}^{-1}Q^{2-} &=& -ic_0 \sqrt{\frac{\mu}{3}}\left(
a_0^{a\dagger}\gamma^a S_0 + a_0^a\gamma^aS_0^{\dagger}
\right) 
-i \sum_{n=1}^{\infty}c_n\sqrt{\omega_n}
\left(
a_n^{a\dagger}\gamma^a S_n + a_n^a\gamma^a S_n^{\dagger} 
\right)
\nn \\
&& \quad 
+ \frac{1}{2}\cdot\frac{\mu}{3}\sum_{n=1}^{\infty}\frac{1}{c_n\sqrt{\omega_n}}
\left(
\bar{a}^a_n\gamma^a\Pi\tilde{S}_n^{\dagger} - \bar{a}_n^{a\dagger}\gamma^a
\Pi\tilde{S}_n
\right)\,, 
\end{eqnarray}
and 
\begin{eqnarray}
\sqrt{\al'}^{-1}Q'{}^{2-} &=& 
c_0' \sqrt{\frac{\mu}{6}} \left(a_0^{b\dagger}\gamma^a\Pi^{\s T}S_0' 
- a_0^b\gamma^a\Pi^{\s T}S_0'{}^{\dagger}\right) 
-i \sum_{n=1}^{\infty}c_n'\sqrt{\omega_n'}\left(
\bar{a}_n^{b\dagger}\gamma^b\tilde{S}'_n + \bar{a}_n^b \gamma^b 
\tilde{S}_n'{}^{\dagger}
\right)
\nn \\
&& \quad + \frac{1}{2}\cdot\frac{\mu}{6}\sum_{n=1}^{\infty}
\frac{1}{c_n'\sqrt{\omega_n'}}\left(
a_n^{b\dagger}\gamma^b\Pi^{\s T}S'_n 
- a_n^b \gamma^b\Pi^{\s T}S_n'{}^{\dagger}
\right)\,, \\
\sqrt{\al'}^{-1}Q'{}^{1+} &=& -ic_0' \sqrt{\frac{\mu}{6}}\left(
a_0^{b\dagger}\gamma^b S_0' + a_0^b\gamma^bS_0'{}^{\dagger}
\right) 
-i \sum_{n=1}^{\infty}c_n'\sqrt{\omega_n'}
\left(
a_n^{b\dagger}\gamma^b S_n' + a_n^b\gamma^b S_n'{}^{\dagger} 
\right)
\nn \\
&& \quad 
+ \frac{1}{2}\cdot\frac{\mu}{6}\sum_{n=1}^{\infty}
\frac{1}{c_n'\sqrt{\omega_n'}}
\left(
\bar{a}^b_n\gamma^b\Pi\tilde{S}_n'{}^{\dagger} - \bar{a}_n^b
\gamma^b
\Pi\tilde{S}_n'
\right)\,.
\end{eqnarray} 
On the other hand, the kinematical supersymmetries are rewritten as 
\begin{eqnarray}
&& \tilde{Q}^{1-} = i\Pi\tilde{\Psi}_0 = \frac{\Pi}{2}(S_0 - S_0^{\dagger})\,,
\quad \tilde{Q}^{2+} = i\Psi_0 
=  \frac{i}{2}\left(S_0 + S_0^{\dagger}\right)\,, \\ 
&& \tilde{Q}^{1+} = i\Pi\tilde{\Psi}_0 = \frac{\Pi}{2}(S_0 - S_0^{\dagger})\,,
\quad \tilde{Q}^{2-} = i\Psi_0 
=  \frac{i}{2}\left(S_0 + S_0^{\dagger}\right)\,. 
\end{eqnarray}

In the next section, the above expressions of supercharges will be used 
for constructing the fermionic boundary states. 

\section{Boundary States of Type IIA String Theory}

In this section we will construct the boundary states of type IIA string
theory on the pp-wave background with 24 supersymmetries. 
To begin with, the bosonic boundary states will be constructed. 
Next, we construct the fermionic boundary states and classify the 
allowed D-brane instantons in our theory. 
The resulting boundary states will be used for the calculation of 
amplitude in the closed string description.

\subsection{Bosonic Boundary States}

Here we will consider the bosonic part of boundary states in the Type
IIA string theory. The bosonic coordinates $(x^a,x^b)~~(a=1,\ldots,4, 
b=5,\ldots,8)$ are classified into $(x^{\bar{a}},x^{\bar{b}})$ 
(for the Neumann condition) 
and $(x^{\underline{a}},x^{\underline{b}})$ 
(for the Dirichlet condition). 

The definition of bosonic boundary state $|B\ra$ is 
given by the following boundary conditions: 
\begin{eqnarray}
\label{nb}
\partial_{\tau}x^{\bar{a}}|_{\tau =0} |B\ra = 0\,, 
\quad  \partial_{\tau}x^{\bar{b}}|_{\tau =0} |B\ra = 0 
\quad &&(\mbox{Neumann})\,, \\
(x_0^{\underline{a}} - q_0^{\underline{a}})|_{\tau =0} | B \ra = 0\,, 
\quad  
(x_0^{\underline{b}} - q_0^{\underline{b}})|_{\tau =0} | B \ra = 0 
\quad && (\mbox{Dirichlet})\,. 
\label{db}
\end{eqnarray}
The conditions (\ref{nb}) can be rewritten as 
\begin{eqnarray}
(a_0^{\bar{a}} + a_0^{\bar{a}\dagger})|B\ra = 0\,, 
\quad (a_n^{\bar{a}} + \bar{a}_n^{\bar{a}\dagger})|B\ra = 0 
\quad (n>0)\,, \\
(a_0^{\bar{b}} + a_0^{\bar{b}\dagger})|B\ra = 0\,, 
\quad (a_n^{\bar{b}} + \bar{a}_n^{\bar{b}\dagger})|B\ra = 0 
\quad (n>0)\,,
\end{eqnarray}
and lead to the bosonic boundary state $|B\ra$ 
for Neumann directions described by 
\begin{eqnarray}
| B \ra &=& \exp\left(-\frac{1}{2}\sum_{\bar{a}}a_0^{\bar{a}\dagger}
a_0^{\bar{a}\dagger} - \frac{1}{2}\sum_{\bar{b}}a_0^{\bar{b}\dagger}
a_0^{\bar{b}\dagger} - \sum_{n=1}^{\infty}
\left\{
\sum_{\bar{a}}a_n^{\bar{a}\dagger}\bar{a}_n^{\bar{a}\dagger} 
+ \sum_{\bar{b}}a_n^{\bar{b}\dagger}\bar{a}_n^{\bar{b}\dagger} 
\right\}\right) |0\ra \,,
\end{eqnarray}
where $|0\ra$ is the bosonic Fock vacuum state annihilated by the operators,  
$a_n^{a,b}$ and $\bar{a}_n^{a,b}$. 

On the other hand, the second conditions (\ref{db}) 
can be rewritten as 
\begin{eqnarray}
\left(a_0^{\underline{a}} - a_0^{\underline{a}\dagger} 
+ i\left(\frac{2\mu}{3\al'}\right)^{1/2}q_0^{\underline{a}}\right) 
| B\ra = 0\,, 
\quad (a_n^{\underline{a}} - \bar{a}_n^{\underline{a}\dagger}) | B\ra =0\,, \\
\left(a_0^{\underline{b}} - a_0^{\underline{b}\dagger} 
+ i\left(\frac{\mu}{3\al'}\right)^{1/2}q_0^{\underline{b}}\right) 
| B\ra = 0\,, 
\quad (a_n^{\underline{b}} - \bar{a}_n^{\underline{b}\dagger}) | B\ra =0\,.
\end{eqnarray}
With these boundary conditions, we can construct the boundary states for
Dirichlet directions described by
\begin{eqnarray}
| B \ra &=& \e^{+\frac{1}{2}\sum_{\underline{a}}\left\{
a_0^{\underline{a}\dagger} - i\left(\frac{2\mu}{3\al'}\right)^{1/2}q_0^{\underline{a}}
\right\}^2 
+ \frac{1}{2}\sum_{\underline{b}}
\left\{
a_0^{\underline{b}\dagger} -i \left(\frac{\mu}{3\al'}\right)^{1/2}q_0^{\underline{b}}
\right\}^2
}\cdot \e^{+\sum_{n=1}^{\infty}\left\{
\sum_{\underline{a}}a_n^{\underline{a}\dagger}\bar{a}_n^{\underline{a}\dagger}
+ \sum_{\underline{b}}a_n^{\underline{b}\dagger}
\bar{a}_n^{\underline{b}\dagger}
\right\}}|0\ra\,.
\end{eqnarray}
Here we shall introduce a diagonal matrix $M_{ij}=\mbox{diag}(\pm
1,\cdots,\pm 1)$ with eight components where $+1$ is assigned for
Dirichlet directions and $-1$ is assigned for Neumann ones. If we will
set as $q_0^{\underline{a}}=q_0^{\underline{b}}=0$, the bosonic boundary
state can be rewritten as
\begin{eqnarray} 
\label{bvs}
| B \ra &=& | B \ra_{\mu/3}\otimes | B \ra_{\mu/6} \,, \\
| B \ra_{\mu/3} &\equiv& \e^{+ \frac{1}{2}M_{aa'}
a_0^{a\dagger}a_0^{a'\dagger}}\cdot 
\e^{\sum_{n=1}^{\infty}M_{aa'}a_n^{a\dagger}\bar{a}_n^{a'\dagger}
}|0\ra\,, \nn \\
| B \ra_{\mu/6} &\equiv& \e^{+ \frac{1}{2}M_{bb'}
a_0^{b\dagger}a_0^{b'\dagger}}\cdot 
\e^{\sum_{n=1}^{\infty}M_{bb'}a_n^{b\dagger}\bar{a}_n^{b'\dagger}
}|0\ra\,. \nn 
\end{eqnarray}
Thus the bosonic boundary state is the product of 
two sectors with mass $\mu/3$ and that with $\mu/6$, and 
it has the $SO(4)\times SO(4)$ symmetry. 
We will study the fermionic boundary states in the next subsection. 

\subsection{Fermionic Boundary States}

We will now consider the fermionic part of 
boundary states in our case. The fermionic boundary states 
are defined by 
\begin{eqnarray}
\label{orig}
&&\left(Q_{\al}^{2-} - i\eta M_{\al\beta}^{(\mu/3)}
Q_{\beta}^{1+}\right) 
| B \ra = 0\,, \quad \left(
Q'_{\al}{}^{2-} - i\eta M_{\al\beta}^{(\mu/6)}Q'_{\beta}{}^{1+}
\right)| B \ra = 0 \,, \\
&& \left(
\tilde{Q}_{\al}^{2-} + i\eta \hat{M}_{\al\beta}^{(\mu/3)}
\tilde{Q}^{1+}_{\beta}
\right) | B \ra = 0\,, \quad 
\left(
\tilde{Q}^{2+} + i\eta 
\hat{M}_{\al\beta}^{(\mu/6)}\tilde{Q}^{1-}_{\beta}
\right) | B \ra = 0\,,
\label{kinematical}
\end{eqnarray}
where matrices $M_{\al\beta}^{(\mu/3)}$,  
$M^{(\mu/6)}_{\al\beta}$,  $\hat{M}_{\al\beta}^{(\mu/3)}$
and $\hat{M}^{(\mu/6)}_{\al\beta}$
satisfy the following relations:
\begin{eqnarray}
&& M_{\al\beta}(M^{\s T})_{\beta\gamma} = \del_{\al\gamma}\,, 
\quad (M^{\s T})_{\al\beta}M_{\beta\gamma} 
= \del_{\al\gamma}\,, \\ 
&& \hat{M}_{\al\beta}(\hat{M}^{\s T})_{\beta\gamma} 
= \del_{\al\gamma}\,, 
\quad (\hat{M}^{\s T})_{\al\beta}\hat{M}_{\beta\gamma} 
= \del_{\al\gamma}\,.
\end{eqnarray}
The definition of the fermionic boundary states (\ref{kinematical})
leads to the conditions written in terms of the zero-modes\footnote{If
we redefine the $\widetilde{\Psi}_0$ as $\Pi\widetilde{\Psi}_0
\rightarrow \widetilde{\Psi}_0$, then we obtain the usual expressions
for zero-mode conditions. The effect from the redefinition of
$\widetilde{\Psi}_0$ are absorbed into the definition of the creation
and annihilation operators without the modification of anticommutation
relations, and so we have no trouble for our discussion.}:
\begin{eqnarray}
\label{fvs}
\left((\Psi_0)_{\al} + i\eta \hat{M}^{(\mu/3)}_{\al\beta}
(\Pi\widetilde{\Psi}_0)_{\beta}
\right) | B \ra = 0\,, \quad 
\left((\Psi_0')_{\al} + i\eta \hat{M}^{(\mu/6)}_{\al\beta}
(\Pi\widetilde{\Psi}_0')_{\beta}
\right) | B \ra = 0\,.
\end{eqnarray}
These conditions suggest us to take the following ansatz: 
\begin{eqnarray}
\left(
(S_n)_{\al} 
+ i\eta \hat{M}^{(\mu/3)}_{\al\beta}
(\tilde{S}_n^{\dagger})_{\beta}
\right)|B\ra = 0\,, \quad 
\left(
(S'_n)_{\al}  
+ i\eta \hat{M}^{(\mu/6)}_{\al\beta}
(\tilde{S}'_n{}^{\dagger})_{\beta}
\right) | B \ra = 0\,.
\label{4.18}
\end{eqnarray}
The above equations can be easily solved and 
the boundary state is given by 
\begin{eqnarray}
|B\ra = \e^{\sum_{n=1}^{\infty}\left\{
-i\eta \hat{M}^{(\mu/3)}_{\al\beta}(S_n^{\dagger})_{\al}
(\tilde{S}_n^{\dagger})_{\beta} 
-i\eta \hat{M}^{(\mu/6)}_{\al\beta}(S_n'{}^{\dagger})_{\al}
(\tilde{S}_n'{}^{\dagger})_{\beta}
\right\}
}|B\ra_0\,,  
\label{sol}
\end{eqnarray}
where $|B\ra_0$ is the fermionic vacuum state yet to be determined. 
By the way, this boundary state is by definition the state satisfying
(\ref{orig}) and (\ref{kinematical}). If we now act the conditions
(\ref{4.18}) on (\ref{orig}), we have three types of conditions 
that lead us to determine the structure of the matrices $M^{(\mu/3)}$,
$M^{(\mu/6)}$, $\hat{M}^{(\mu/3)}$ and $\hat{M}^{(\mu/6)}$. 
Firstly, we obtain the conditions 
\begin{eqnarray}
M^{a'a}\gamma^{a'} = -M^{(\mu/3)}\gamma^a \hat{M}^{(\mu/3){\s T}}\,, \quad 
M^{b'b}\gamma^{b'} = -M^{(\mu/6)}\gamma^b \hat{M}^{(\mu/6){\s T}}\,,
\label{cd1}
\end{eqnarray}
which are similar to those arising in flat space \cite{Green}.  The
second type of conditions, which appears only in the pp-wave case, is
\begin{eqnarray}
M^{aa'}\gamma^{a'}\Pi = - M^{(\mu/3)}\gamma^a\Pi \hat{M}^{(\mu/3)}\,, \quad
M^{bb'}\gamma^{b'}\Pi = - M^{(\mu/6)}\gamma^b\Pi \hat{M}^{(\mu/6)}\,.
\label{cd2}
\end{eqnarray}
Finally, the third type of conditions we get comes from the zero-mode 
parts: 
\begin{eqnarray}
\label{zero}
&& \left\{a_0^{a\dagger}\left(
\gamma^a + \eta M^{(\mu/3)}\gamma^a\Pi
\right) S_0 
+M^{aa'}a_0^{a'\dagger}\left(
\gamma^a - \eta M^{(\mu/3)}\gamma^a\Pi 
\right) S_0^{\dagger}\right\} | B \ra = 0\,, \\
&& \left\{a_0^{b\dagger}\left(
\gamma^b + \eta M^{(\mu/6)}\gamma^b\Pi
\right) S_0 
+M^{bb'}a_0^{b'\dagger}\left(
\gamma^b - \eta M^{(\mu/6)}\gamma^b\Pi 
\right) S_0^{\dagger}\right\} | B \ra = 0\,.
\label{zero2}
\end{eqnarray}
By the use of the definition of fermionic vacuum $S_0 | B\ra_0 =0$ and 
the identities $a_0^a = M^{aa'} a_0^{a'\dagger}$ 
and $a_0^b = M^{bb'}a_0^{b'\dagger}$, we can rewrite 
(\ref{zero}) and (\ref{zero2}) as 
\begin{eqnarray}
\label{co}
&&\left\{
p_0^{a'}\left(
\del^{aa'} - M^{aa'}
\right)
+ i \frac{\mu}{3\al'}x_0^{a'}
\left(\del^{aa'} + M^{aa'}\right)
\right\}
\left(
\gamma^a - \eta M^{(\mu/3)}\gamma^a\Pi 
\right) 
S_0^{\dagger} | B \ra_0 = 0\,, \\
&&\left\{
p_0^{b'}\left(
\del^{bb'} - M^{bb'}
\right)
+ i \frac{\mu}{6\al'}x_0^{b'}
\left(\del^{bb'} + M^{bb'}\right)
\right\}
\left(
\gamma^b - \eta M^{(\mu/6)}\gamma^b\Pi 
\right) 
S_0^{\dagger} | B \ra_0 = 0\,.
\label{co2}
\end{eqnarray}
Using the conditions (\ref{co}) and (\ref{co2}), we can read off
supersymmetries preserved by D-brane instantons in terms of their
positions. In the case that all position coordinates of a D-brane
instanton $q^{r}$ for the Dirichlet directions equals zero, the D-brane
instanton has 12 (4 dynamical + 8 kinematical) supersymmetries (i.e.,
1/2(=12/24) BPS D-brane instanton).  If the position coordinates for the
Dirichlet directions are not at the origin, then the D-brane instanton
has 8 (0+8) supersymmetries (i.e., 1/3(=8/24) BPS D-brane instanton).
Thus all of the dynamical supersymmetries are broken.  However the
D-brane instantons apart from the origin are supersymmetric solutions
since they have 8 kinematical supersymmetries.

As a final remark regarding the structure of the matrices,
$M^{(\mu/3)}$, $M^{(\mu/6)}$, $\hat{M}^{(\mu/3)}$ and
$\hat{M}^{(\mu/6)}$, we now consider the chirality condition while the
above three types of conditions are almost same as in the type IIB
string case \cite{BP}. First we can easily find that both matrices
$M^{(\mu/3)}$ and $M^{(\mu/6)}$ contain the odd number of gamma matrices
in order to preserve the $SO(8)$ chirality measured by
$\gamma^9$. Moreover, if we consider the chirality in terms of the
matrix $R = \gamma^{1234}$, we have the following conditions basically
from (\ref{orig}) and (\ref{kinematical}):
\begin{eqnarray}
\{R,\,M^{(\mu/3)}\} = 0\,, \quad 
\{R,\,M^{(\mu/6)}\} = 0\,, \quad \{R,\,\hat{M}^{(\mu/3)}\} = 0\,, \quad 
\{R,\,\hat{M}^{(\mu/6)}\} = 0\,.
\label{cd4}
\end{eqnarray}
Then we obtain the following complete boundary state: 
\begin{eqnarray}
|B\ra &=& 
\e^{\sum_{n=1}^{\infty}\left\{M_{aa'}a_n^{a\dagger}\bar{a}_n^{a'\dagger}
+ M_{bb'}a_n^{b\dagger}\bar{a}_n^{b'\dagger}   
-i\eta \hat{M}^{(\mu/3)}_{\al\beta}(S_n^{\dagger})_{\al}
(\tilde{S}_n^{\dagger})_{\beta} 
-i\eta \hat{M}^{(\mu/6)}_{\al\beta}(S_n'{}^{\dagger})_{\al}
(\tilde{S}_n'{}^{\dagger})_{\beta}
\right\}
} | B\ra_0  \nn \\
|B\ra_0 &=& \left(
M_{\s IJ}| I \ra | J \ra - i \eta M_{\al\beta}|\al\ra | \beta\ra
\right)
e^{+ \frac{1}{2}M_{aa'}
a_0^{a\dagger}a_0^{a'\dagger} + \frac{1}{2}M_{bb'}
a_0^{b\dagger}a_0^{b'\dagger}} | 0\ra \,,
\end{eqnarray}
where the state $|B\ra_0$ is the product of the bosonic vacuum state,
which is given by picking up the zero-mode parts in Eq.\,(\ref{bvs}),
and the fermionic one which is the solution of (\ref{fvs}).

The remaining task is to determine the matrices $M$ and $\hat{M}$ 
from the conditions (\ref{cd1}), (\ref{cd2}), (\ref{co}), 
(\ref{co2}) and (\ref{cd4}). 
The determined structure of the matrices leads to the classification 
of possible D-brane instantons. 
This will be done in the next subsection.

\subsection{Classification of D-brane Instantons}

We will classify the allowed D-brane instantons by determining 
the matrices $M$ and $\hat{M}$.  
Now let us analyze each of D$p$-brane instantons. 
\begin{itemize}
\item[\bf D0:] D0-brane instantons are expressed by the following matrices: 
\begin{eqnarray}
M^{(\mu/3)} = M^{(\mu/6)} = \hat{M}^{(\mu/3)} = \hat{M}^{(\mu/6)}
= \gamma^{\s I} ~~~(I=1,2,3)\,.
\end{eqnarray} 
The $x^{\s I}$-direction satisfies the Neumann boundary condition and 
other directions satisfy the Dirichlet boundary condition. 
When we consider $M=\gamma^1$ as an example, $x^1$ is a Neumann 
direction and other directions are Dirichlet ones. 
If we consider the D0-brane instanton at the origin 
$q^{2,3,4,5,6,7,8}=0$, then it is a 1/2 BPS object. 
If we consider the D0-brane instanton apart from the origin, then 
it becomes a 1/3 BPS object.

\item[\bf D2:] D2-brane instantons are given by 
\begin{eqnarray}
&& M^{(\mu/3)} = M^{(\mu/6)} = \hat{M}^{(\mu/3)} = \hat{M}^{(\mu/6)}= \gamma^{{\s IJ}4}\,, \quad \\
\mbox{or}& & \quad 
\quad M^{(\mu/3)} = M^{(\mu/6)} = \hat{M}^{(\mu/3)} = \hat{M}^{(\mu/6)} = 
\gamma^4\gamma^{bb'}\,, \nn
\end{eqnarray}
where $I$ and $J$ take values in 1,\,2,\,3, and $b$ and $b'$ run from 
5 to 8. For example, if we take $M=\gamma^{124}$ then 
$x^{1,2,4}$-directions satisfy the Neumann condition and 
others are Dirichlet directions. When the D2-brane instanton sits 
at the origin $q^{3,5,6,7,8}=0$, it is a 1/2 BPS object. 
Once it goes away from the origin, it becomes 1/3 BPS. 

\item[\bf D4:] D4-brane instantons are described by 
\begin{eqnarray}
&& M^{(\mu/3)} = M^{(\mu/6)} =  \hat{M}^{(\mu/3)} = \hat{M}^{(\mu/6)}
=\gamma^{123}\gamma^{bb'}\,, \quad \\
\mbox{or}& & \quad 
M^{(\mu/3)} = M^{(\mu/6)}= \hat{M}^{(\mu/3)} = \hat{M}^{(\mu/6)} = \gamma^{\s I}\gamma^{5678}\,. \nn 
\end{eqnarray}
When we take $M=\gamma^{12356}$, the $x^{1,2,3,5,6}$-directions 
satisfy the Neumann boundary condition 
and others are the Dirichlet directions.  
When the D4-brane instanton is at the origin $q^{4,7,8}=0$, 
it is a 1/2 BPS object. If it is apart from the origin, 
it becomes 1/3 BPS.

\item[\bf D6:] D6-brane instantons are given by 
\begin{eqnarray}
M^{(\mu/3)} = M^{(\mu/6)}= \hat{M}^{(\mu/3)} = \hat{M}^{(\mu/6)} = \gamma^{{\s IJ}4}\gamma^{5678}\,.
\end{eqnarray}
For example, the case $M=\gamma^{1245678}$ leads to a D6-brane 
instanton that preserves 12 supersymmetries (i.e., 1/2 BPS) 
for $q^3=0$ and 8 supersymmetries (1/3 BPS) for $q^3 \neq 0$.  
\end{itemize}

We should remark that the above classification of D-brane instantons 
is consistent with that of D-branes in the open string description 
\cite{HS3,HPS}. 
The Neumann (Dirichlet) boundary condition in the closed string
description is simply related by the Dirichlet (Neumann) one in the open
string description, and hence this identification should hold as a
matter of course.  For comparison with the classification 
of D-branes \cite{HS3,HPS}, 
we shall summarize our result in {\bf Tab.}~\ref{list:tab}. 

\begin{table}
\begin{center}
 \begin{tabular}{c|c}
\hline
$N_N$ &  $M^{(\mu/3)}=M^{(\mu/6)} = \hat{M}^{(\mu/3)} = 
\hat{M}^{(\mu/6)} $ \\
\hline 
\hline 
1  &  $\gamma^{\s I}$ \\ 
3 & \quad $\gamma^{{\s IJ}4}$\,, \quad
  $\gamma^{4}\gamma^{bb'}$ \quad \\
5 & \quad $\gamma^{123}\gamma^{bb'}$\,, \quad $\gamma^{\s I}\gamma^{5678}$ 
\quad  \\
7 & $\gamma^{{\s IJ}4}\gamma^{5678}$ \\
\hline
 \end{tabular}
\end{center}
\caption{\footnotesize List of possible 
D-brane instantons in our Type IIA string
 theory. The $N_N$ is the number of Neumann directions.} \label{list:tab}
\end{table}

\section{Cylinder Amplitude in the Closed String Description}

In this section we will calculate the tree amplitude in the closed string
description. The interaction energy between a pair of D-branes comes
from the exchange of a closed string between two boundary states
(i.e., a cylinder diagram). 

The expression of the cylinder diagram (tree diagram) 
in the light-cone formulation can be expressed as 
\begin{eqnarray}
&&\mathcal{A}_{Dp_1;Dp_2}(x^+,x^-,q_1^i,q_2^j) \\
&& = \int\!\!\frac{dp^+dp^-}{2\pi i}\,\e^{ip^+x^- + ip^-x^+} 
\la Dp_1,-p^-,-p^+,q_1^i |\left(
\frac{1}{p^+(p^- + H)}
\right)
| Dp_2,p^-,p^+,q_2^j \ra \nn \\
&& = \int^{+\infty}_{-\infty}\!\!dp^+\,
\e^{ip^+x^-} \frac{\th(p^+)}{p^+}
\la Dp_1,-p^+,q_1^i | 
\,\e^{-iHx^+}  | Dp_2,p^+,q_2^j\ra \,, \nn 
\end{eqnarray}
where $H$ is the light-cone Hamiltonian of a closed string and the $|
Dp,p^+,q^i \ra$ represents a boundary state of a D$p$-brane instanton
located at the transverse position $q^i$ with the longitudinal momentum
$p^+$.  We note that the prescription given in \cite{BGG} has been used
for obtaining the last line in the above equation.  If we define the
variable $t$ by 
\[
 x^+ = \pi \tau = -i\pi t 
\]
by performing the customary Wick rotation, the amplitude is
rewritten as 
\begin{eqnarray}
\mathcal{A}_{Dp_1;Dp_2}(x^+,x^-,q_1^i,q_2^j) = 
\int^{\infty}_0\!\!\!\! dt\,\e^{- \frac{x^+x^-}{\pi t}}
\tilde{\mathcal{A}}_{Dp_1;Dp_2}(t,q_1^i,q_2^j)\,, 
\end{eqnarray}
where $\tilde{A}_{Dp_1;Dp_2}(t,q_1,q_2)$ 
is the expectation value: 
\begin{eqnarray}
\label{tree}
\tilde{\mathcal{A}}_{Dp_1;Dp_2}(t,q_1^i,q_2^j) \equiv 
\la Dp_1,-p^+,q_1^i | \,\e^{-2\pi t (H/2) }
| Dp_2,p^+,q_2^j \ra \,.
\end{eqnarray}
 
Now the tree amplitude (\ref{tree}) will be calculated by using 
the boundary states constructed before. We restrict ourselves 
to the case of identical D$p$-brane
instantons for simplicity.  
The calculus consists of three parts: 1) vacuum energies, 
2) nonzero-modes, and 3) zero-modes.  

Let us concentrate only on the sector with mass $\nu\equiv\mu/3$.  The
other sector with mass $\nu'\equiv \mu/6$ results in the same final
expression only with the difference in the value of mass parameter.
When we consider the contribution of vacuum energies to the amplitude,
the evaluation of vacuum energies is identical to that of the partition
function in the closed string case:
\begin{eqnarray}
\e^{2\Delta(\nu;0)} \quad (\mbox{for bosons})\,, 
\quad \e^{-2\Delta(\nu;0)} \quad (\mbox{for fermions})\,.
\end{eqnarray} 
The contribution of nonzero-modes to the amplitude is also the same as
that of partition function of a closed string, and so it is
readily written as
\begin{eqnarray}
\prod_{n=1}^{\infty}
\left(1 - q^{\omega_n}\right)^{-4} \quad (\mbox{bosons})\,, \quad 
\prod_{n=1}^{\infty}
\left(1 - q^{\omega_n}\right)^{4} \quad (\mbox{fermions})\,, 
\quad q\equiv e^{-2\pi t}\,.
\end{eqnarray} 
The contribution of bosonic zero-modes can be evaluated 
by using the formula: 
\begin{eqnarray}
\la 0 | \e^{\pm\frac{1}{2}a_0a_0}q^{\pm \frac{1}{2}m a_0^{\dagger}a_0}
\e^{\pm \frac{1}{2}a_0^{\dagger}a_0^{\dagger}} | 0 \ra 
= \left(1-q^m\right)^{-1/2}\,.
\end{eqnarray}
As a result, the factor $(1-q^{\nu})^{-2}$ is obtained from bosonic 
zero-modes. The fermionic zero-modes can be evaluated by adopting the
prescription given in the appendix of \cite{BGG}, and the resulting
contribution is $(1-q^{\nu})^2$. Consequently, the contribution of
zero-modes is summarized as follows:  
\begin{eqnarray}
(1-q^{\nu})^{-2} \quad (\mbox{for bosons})\,, \quad 
(1-q^{\nu})^{2} \quad (\mbox{for fermions})\,.
\end{eqnarray} 

After taking account of the sector with mass $\mu/6$, the total
partition function is then represented by
\begin{eqnarray}
\tilde{A}_{Dp;Dp} &=& \tilde{A}_{Dp;Dp}^{B}\cdot \tilde{A}_{Dp;Dp}^{F} 
\;=\; \mathcal{N}^2_{Dp}\,,
\label{5.8}
\end{eqnarray} 
where $\mathcal{N}_{Dp}$ is the normalization factor of boundary states,
which is not determined yet. This factor can be fixed by calculating the
cylinder diagram in the open string channel.  This task will be done in
the next section.
 
In the work of \cite{BGG}, the ``conformal field theory condition''
\begin{eqnarray}
\label{CFT}
\tilde{\mathcal{A}}_{Dp_1;Dp_2}(t,q_1,q_2) 
= \tilde{Z}_{Dp_1;Dp_2}(\tilde{t},q_1,q_2)\,, 
\end{eqnarray}
was analyzed in the case of the pp-wave background.  This condition
ensures the consistency between closed and open string channels.  We now
turn to the calculation of the partition function with D-branes (i.e.,
the cylinder diagram) in order to show that the condition (\ref{CFT})
also holds in our theory.

\section{Partition Function in the Open String Description}

In this section we will discuss the one-loop amplitude of 
open string, and confirm the consistency condition between open and
closed string channels. 

We start from the light-cone 
action of open string defined by 
\begin{eqnarray}
S_{\rm open} &=& \frac{1}{4\pi\al'}\int\!\! d\tau\!
\int^{\pi}_0\!\!\! d\sigma\, 
\Biggl[
\sum^8_{i=1}
\left\{
(\partial_{\tau}x^i)^2-(\partial_{\sigma}x^i)^2
\right\}
-\left(\frac{\mu}{3}\right)^2\sum^4_{a=1}(x^a)^2
-\left(\frac{\mu}{6}\right)^2\sum^8_{b=5}(x^b)^2
\Biggr] \nn \\
&&  + 
 \frac{i}{2\pi}\int\!\! d\tau\!
\int^{\pi}_0\!\!\! d\sigma\, 
\Biggl[
\Psi^{1+{\s T}}\partial_{-}\Psi^{1+}+
\Psi^{1-{\s T}}\partial_{-}\Psi^{1-}+
\Psi^{2+{\s T}}\partial_{+}\Psi^{2+}+
\Psi^{2-{\s T}}\partial_{+}\Psi^{2-}\nn \\
&& 
-\frac{\mu}{3}\Psi^{1-{\s T}}\Pi^{\s T}\Psi^{2+}
-\frac{\mu}{6}\Psi^{1+{\s T}}\Pi^{\s T}\Psi^{2-}
+\frac{\mu}{6}\Psi^{2-{\s T}}\Pi\Psi^{1+}
+\frac{\mu}{3}\Psi^{2+{\s T}}\Pi\Psi^{1-}\Biggr]\,.  
\end{eqnarray}

To begin with, we shall present the mode-expansion in the case of open string. 
The mode-expansion of D-D string is 
expressed as  
\begin{eqnarray}
x^a(\tau,\sigma) &=& q^a_0\cdot \frac{\sinh\nu (\pi -\sigma)}{\sinh (\pi\nu)}
+ q^a_1\cdot \frac{\sinh (\nu\sigma)}{\sinh (\pi\nu)} -\sqrt{2\alpha'}
\sum_{n\neq 0}\frac{1}{\omega_n}\alpha^a_ne^{-i\omega_n\tau}
\sin (n\sigma)\,, \\
x^b(\tau,\sigma) &=& q^b_0\cdot \frac{\sinh\nu' (\pi -\sigma)}{\sinh (\pi\nu')}
+ q^b_1\cdot \frac{\sinh (\nu'\sigma)}{\sinh (\pi\nu')} -\sqrt{2\alpha'}
\sum_{n\neq 0}\frac{1}{\omega'_n}\alpha^a_ne^{-i\omega'_n\tau}
\sin (n\sigma)\,, 
\end{eqnarray}
where the endpoints satisfy the Dirichlet conditions: 
$x^i(\sigma = 0) = q^i_0\,,\,\,x^i(\sigma = \pi) = q^i_1$.

The mode expansion of N-N string,  
whose endpoints satisfy the Neumann conditions, 
is written as 
\begin{eqnarray}
x^a(\tau,\sigma) &=&x^a_0\cos (\nu\tau)
+\frac{1}{\nu}\cdot 2\alpha' p^a_0\sin (\nu\tau) 
+ i\sqrt{2\alpha'}\sum_{n\neq 0}
\frac{1}{\omega_n}
\alpha^a_n e^{-i\omega_n \tau}
\cos (n\sigma)\,, \\
x^b(\tau,\sigma) &=& x^b_0\cos (\nu'\tau)
+\frac{1}{\nu'}\cdot 2\alpha' p^b_0\sin (\nu'\tau) 
+ i\sqrt{2\alpha'}\sum_{n\neq 0}
\frac{1}{\omega'_n}
\alpha^b_n e^{-i\omega'_n \tau}
\cos (n\sigma)\,.
\end{eqnarray}

The mode-expansion of fermions are the same with that in the case of closed
string, but we have to take account of boundary conditions at $\sigma
=0$ and $\pi$ described by 
\begin{eqnarray}
&&\Psi^{1-} = \Omega\Psi^{2+}\,,\quad 
\Psi^{2+} = \Omega^{\s T}\Psi^{1-}\,,\quad 
\Psi^{1+} = \Omega\Psi^{2-}\,,\quad 
\Psi^{2-} = \Omega^{\s T}\Psi^{1+}\,,\quad \nn \\
&&\tilde{\Psi}_n = \Omega\Psi_n\,,\quad 
\tilde{\Psi}_0 = - \Pi\Omega\Psi_0\,,\quad 
\Psi_0 = \Pi\Omega\tilde{\Psi}_0\,,\quad \Pi\Omega\Pi\Omega = -1\,, \nn 
\end{eqnarray}
where $\Omega$ is the gluing matrix for fermionic modes on the 
boundaries. 

We now introduce the creation and annihilation operators given by  
\begin{eqnarray}
&&a^a_n=\frac{1}{\sqrt{\omega_n}}\alpha^a_n\,,\quad 
a^{a\dagger}_n=\frac{1}{\sqrt{\omega_n}}\alpha^a_{-n}\,, \quad
 [a^a_m,a^{a'\dagger}_n]=\delta^{a,a'}\delta_{m,n} \quad 
(m,n>0)\,,\nn \\
&&a^b_n=\frac{1}{\sqrt{\omega_n'}}\alpha^b_n\,,\quad 
a^{b\dagger}_n = \frac{1}{\sqrt{\omega_n'}}\alpha^b_{-n}\,, \quad
 [a^b_m,a^{b'\dagger}_n]=\delta^{b,b'}\delta_{m,n} \quad 
(m,n>0)\,,\nn \\
&& a^a_0 = \sqrt{\frac{\al'}{\nu}}
\left(p^a_0-i\nu\frac{1}{2\al'}x^a_0\right)\,,\quad 
a^{a\dagger}_0 = \sqrt{\frac{\al'}{\nu}} 
\left(p^a_0+i\nu\frac{1}{2\al'}x^a_0\right)\,,\quad 
[a^a_0,a^{a'\dagger}_0]=\delta^{a,a'}\,, \nn \\
&& a^b_0 = \sqrt{\frac{\al'}{\nu'}}
\left(p^b_0-i\nu'\frac{1}{2\al'}x^b_0\right)\,,\quad 
a^{b\dagger}_0 = \sqrt{\frac{\al'}{\nu'}} 
\left(p^b_0+i\nu'\frac{1}{2\al'}x^b_0\right)\,,\quad 
[a^b_0,a^{b'\dagger}_0]=\delta^{b,b'}\,,\nn 
\end{eqnarray}
then the Hamiltonian $H_B$ for the Dirichlet directions is expressed by 
\begin{eqnarray}
H_B &=&\frac{1}{4\pi \alpha'}
\cdot \frac{\nu}{\sinh (\pi\nu)}
\left[
\{(q^a_0)^2+(q^a_1)^2\}\cosh (\pi\nu)-2q^a_0 q^a_1
\right] \nn \\
&& + \frac{1}{4\pi \alpha'}
\cdot \frac{\nu'}{\sinh (\pi\nu')}
\left[
\{(q^b_0)^2+(q^b_1)^2\}\cosh (\pi\nu') - 2q^b_0 q^b_1
\right] \nn \\
&& +\frac{1}{2}\sum_{n=1}^{\infty}\omega_n
(a^{a\dagger}_n a^a_n + a^{a\dagger}_n a^a_n)
+\frac{1}{2}\sum_{n=1}^{\infty}\omega'_n
(a^{b\dagger}_n a^b_n + a^{b\dagger}_n a^b_n)\,,
\end{eqnarray}
and that for the Neumann directions is represented by 
\begin{eqnarray}
H_B &=&\frac{1}{2}\omega_0(a^{\dagger\,a}_0a^a_0+a^a_0a^{\dagger\,a}_0)
+\frac{1}{2}\sum_{n\geq 1}\omega_n
(a^{\dagger\,a}_na^a_n+a^a_na^{\dagger\,a}_n) \nn \\
&& +\frac{1}{2}\omega'_0(a^{b\dagger}_0 a^b_0+a^b_0a^{b\dagger}_0)
+\frac{1}{2}\sum_{n\geq 1}\omega'_n
(a^{b\dagger}_n a^b_n + a^b_n a^{b\dagger}_n)\,.
\end{eqnarray}
The Hamiltonian of fermions $H_F$ is rewritten as  
\begin{eqnarray}
H_F &=& \sum_{n=1}^{\infty}(\omega_nS^{\dagger}_nS_n+{\omega'}_n
{S'}^{\dagger}_nS'_n)
-\frac{\mu}{3}i\Psi^{\s T}_0\Pi\Omega\Psi_0
-\frac{\mu}{6}i{\Psi'}^{\s T}_0\Pi\Omega {\Psi'}_0\,. 
\end{eqnarray}
Now we shall evaluate the Casimir energy given as follows: 
\begin{eqnarray}
&& \sum_{n =1}^{\infty}\frac{1}{2}\omega_n\cong
\frac{1}{2}\left(
\sum^{\infty}_{n=1}\sqrt{n^2+\nu^2}-\int^{\infty}_0\!\!dk\,\sqrt{k^2+\nu^2}
\right)
=\frac{1}{2}\left(-\frac{1}{2}\nu+\Delta (\nu ;0)\right)\,, \nn \\
&& \sum_{n =1}^{\infty}\frac{1}{2}\omega_n'\cong
\frac{1}{2}\left(
\sum^{\infty}_{n=1}\sqrt{n^2+\nu'{}^2}-\int^{\infty}_0\!\!dk\,
\sqrt{k^2+\nu'{}^2}
\right)
=\frac{1}{2}\left(-\frac{1}{2}\nu'+\Delta (\nu' ;0)\right)\,, \nn 
\end{eqnarray}
in terms of zero-point energy. 

The zero-point energies of a single boson with the 
Dirichlet condition are given by 
\begin{eqnarray}
&&(D)\,:\,\frac{1}{2}\sum_{n\geq 1}\omega_n
\cong \frac{1}{2}\left(-\frac{1}{2}\nu +\Delta (\nu ;0)\right)
\,, \quad 
\frac{1}{2}\sum_{n\geq 1}\omega'_n
\cong \frac{1}{2}\left(-\frac{1}{2}\nu' +\Delta (\nu' ;0)\right)
\,,\nn 
\end{eqnarray}
and those with the Neumann condition are expressed as  
\begin{eqnarray}
&&(N)\,:\,\frac{1}{2}\sum_{n\geq 0}\omega_n
\cong \frac{1}{2}\left(+\frac{1}{2}\nu +\Delta (\nu ;0)\right)\,, \quad 
\frac{1}{2}\sum_{n\geq 0}\omega'_n
\cong \frac{1}{2}\left(+\frac{1}{2}\nu' +\Delta (\nu' ;0)\right)
\,. \nn 
\end{eqnarray}
As for the zero-point energies for a fermion, we have  
\begin{eqnarray}
E^0 = - 4\cdot \frac{1}{2} \sum_{n\geq 1}\omega_n
\cong \nu - 2\Delta (\nu ;0) 
\,,\quad E'{}^0= - 4 \cdot \frac{1}{2}\sum_{n\geq 1}\omega'_n
\cong \nu' - 2\Delta (\nu' ;0)\,,\nn
\end{eqnarray}
where the factor 4 in front of the summation arises since each of
fermions considered here has four independent components.  The
contributions of zero-point energies to the partition function $Z_F$ are
then
$\e^{-2\pi t (\nu -2\Delta (\nu ;0))}=
\e^{-2\pi t\nu}\e^{4\pi t\Delta (\nu ;0)} 
$ 
and 
$\e^{-2\pi t (\nu' -2\Delta (\nu' ;0))}=
\e^{-2\pi t\nu'}\e^{4\pi t\Delta (\nu' ;0)}\,.$

We note that we have to treat carefully the zero-mode part of the
Hamiltonian
\begin{eqnarray}
H_F^0 = -i\nu \Psi_0^{\s T}\Pi\Omega\Psi_0 
- i\nu' \Psi'_0{}^{\s T}\Pi\Omega\Psi'_0\,. \nn
\end{eqnarray}
The $(\Psi_0)_{\alpha}$ and $(\Psi'_0)_{\alpha}$ have four non-vanishing
components, and hence four sets of creation and annihilation operators
$S^{\pm}_1$, $S^{\pm}_2$ and $S'{}^{\pm}_1$, $S'{}^{\pm}_2$ can be
constructed. Here $S^{+}_{1,2}$ ($S'{}^{+}_{1,2}$) are creation
operators and $S^{-}_{1,2}$ ($S'{}^{-}_{1,2}$) are annihilation ones.
Thus the Hamiltonian can be rewritten as
\begin{eqnarray}
H_F^0 
&=& \frac{\nu}{2} \left(S^{+}_1S^{-}_1 - S^{-}_1S^{+}_1\right)
+\frac{\nu}{2} \left(S^{+}_2S^{-}_2 - S^{-}_2S^{+}_2\right) \nn \\
&& +\frac{\nu'}{2} \left(S'{}^{+}_1S'{}^{-}_1 - S'{}^{-}_1S'{}^{+}_1\right)
+\frac{\nu'}{2} \left(S'{}^{+}_2S'{}^{-}_2 - S'{}^{-}_2S'{}^{+}_2\right)
\,,\nn
\end{eqnarray}
and the associated energies are $\pm \frac{\nu}{2}$ and 
$\pm \frac{\nu'}{2}$\,. Consequently, 
the contribution from these zero-mode parts is evaluated as 
\[
\e^{2\pi \nu t} (1 - \e^{-2\pi \nu t})^2 
+ \e^{2\pi \nu' t} (1 - \e^{-2\pi \nu' t})^2\,.
\]

We now turn to the evaluation of the total partition function of the
open string connecting two identical D$p$-branes.  Let us first consider
the partition function for bosons:
\begin{eqnarray}
Z_B= {\rm Tr}\, \e^{-2\pi tH_B}\,,\quad  q = e^{-2\pi t}\,. \nn
\end{eqnarray}
We will consider the sector with the mass $\nu=\mu/3$. 
The bosonic partition function 
for the ($4-p$) Dirichlet conditions (i.e., $4-p$ = $\sharp$(D-D
strings)) is given by  
\begin{eqnarray}
&&Z_B^{(\nu)} =\frac{1}{\displaystyle \prod_{n\geq 1}(1-q^{\omega_n})^{4-p}}
\cdot q^{(4-p)\cdot \frac{1}{2}\left(
-\frac{1}{2}\nu +\Delta (\nu ;0)
\right)}\cdot f(q^a_0,q^a_1)\,,\nn 
\end{eqnarray}
where the function $f(q_0^a,q_1^a)$ is defined by 
\begin{eqnarray}
&&f(q^a_0,q^a_1) \equiv \exp \left[
\frac{-t}{2\alpha'}
\frac{\nu}{\sinh (\pi\nu)}
\left[
\{(q^a_0)^2+(q^a_1)^2\}\cosh (\pi\nu)
-2q^a_0q^a_1\right]\right]\,. \nn
\end{eqnarray}
The partition function for the $p$ Neumann directions 
(i.e., $p$ = $\sharp$(N-N strings)) is written as 
\begin{eqnarray}
Z_B^{(\nu)}=\frac{1}{\displaystyle \prod_{n\geq 0}(1-q^{\omega_n})^p}
\cdot q^{p\cdot \frac{1}{2}\left(+\frac{1}{2}\nu +\Delta (\nu ;0)\right)}
\,. \nn
\end{eqnarray}
Thus, the bosonic partition function on the sector with
mass $\mu/3$ is represented by  
\begin{eqnarray}
Z_B^{(\nu)} &=& (1 - q^{\nu})^{-p}\cdot 
\frac{1}{\displaystyle \prod_{n\geq 1}(1-q^{\omega_n})^4}
\cdot f(q^a_0,q^a_1)\cdot q^{\nu (-1+\frac{1}{2}p)+2\Delta (\nu ;0)}
\,.\nn \\
 &=& \left(2\sinh (\pi\nu t)\right)^{2-p}
\left[\theta_{(0,0)}(t;\nu)\right]^{-2}\cdot f(q^a_0,q^a_1)\,,\nn
\end{eqnarray}
where we have utilized the theta-like function:
$\theta_{(a,b)}(t;\nu)=
\sqrt{\Theta_{(a,b)}(it;-it;\nu)}$\,.

Next we shall evaluate the partition function for fermions with mass 
$\mu/3$. The fermionic partition function is given by 
\begin{eqnarray}
Z_F = {\rm Tr} (-1)^{\bf F}\e^{-2\pi tH_F}\,.\nn
\end{eqnarray}
After the similar calculation to bosonic case, we obtain 
the fermionic partition function: 
\begin{eqnarray}
Z_F^{(\nu)} &=& (\e^{\pi\nu t}-e^{-\pi\nu t})^2 \cdot 
\e^{-2\pi\nu t}\e^{4\pi t\Delta (\nu ;0)}
\prod_{n=1}^{\infty}(1 - \e^{-2\pi t\omega_n})^4 \nn  \\
&=& \left[\theta_{(0,0)}(t;\nu)\right]^2\,.
\end{eqnarray}

It is an easy task to include the sector with mass $\mu/6$, and thus 
the total partition function is described by 
\begin{eqnarray}
Z_{\rm tot} &=& Z_B^{(\nu)} Z_F^{(\nu)} 
\cdot Z^{(\nu')}_B Z^{(\nu')}_F
  \nn \\
&=& \left(2\sinh (\pi\nu t)\right)^{2-p_1}
\prod_{a\in D}f(q^a_0,q^a_1)\cdot
\left(2\sinh (\pi\nu' t)\right)^{2-p_2} 
\prod_{b\in D}f(q^b_0,q^b_1)\,. 
\label{op-pf} 
\end{eqnarray} 
Here the numbers $p_1$ and $p_2 $ are Neumann directions in the
coordinates $x^a$'s and $x^b$'s, respectively. 
The net number of Neumann directions is represented by 
$\sharp$(Neumann)$=p_1+p_2$. 
The $\prod_{i\in D}$ means the product in terms of 
Dirichlet directions $x^i$'s. 

By comparing the cylinder amplitude (\ref{5.8}) obtained in the last
section with the resulting partition function (\ref{op-pf}) in the case
of $q_0 = q_1 =0$, we can determine the normalization factor of boundary
states.  The modular S-transformation of `massive' theta function
(\ref{theta}) relates the parameter $\mu~(\equiv \mu_{\rm cl})$ in the
closed string to that $\mu~(\equiv \mu_{\rm op})$ in the open string
through the corresponding law: $\mu_{\rm op} t = \mu_{\rm cl}$
\cite{BGG}.  Hence the normalization factor of boundary states for the
D$p$-brane instanton is given by
\begin{equation}
\mathcal{N}_{Dp} = (2\sinh(\pi\mu/3))^{(2-p_1)/2}
\cdot(2\sinh(\pi\mu/6))^{(2-p_2)/2} \quad (p=p_1+p_2)\,.
\end{equation}
Thus, we have shown the open/closed string duality in the
$\mathcal{N}=(4,4)$ type IIA string theory at the origin.  It was
already shown in \cite{BGG} that this duality holds in the case of the
type IIB string theory on the {\it maximally} supersymmetric pp-wave
background.  Although we are in a situation of less supersymmetric case,
the duality still holds at the origin.

It should be noted that the open/closed string duality holds at the
origin. That is, the open/closed string duality requires that there is
no dependence on transverse coordinates since the supersymmetry
conditions require that both D-brane instantons should be at the origin,
as discussed in the paper \cite{BGG}. The cylinder amplitude does not
have sensible behavior once the branes are located away from the origin.

\section{General Properties of Partition Functions of Closed String}

We have discussed the partition function and modular invariance of 
the type IIA string theory on the pp-wave background above. 
In this consideration there are two sectors with masses $\mu/3$ and 
$\mu/6$, 
and we have found that the modular properties hold in each sector. 
In this section, motivated by this fact, 
we will discuss general properties of partition
functions of closed string apart from the type IIA string theory 
considered above. We suggest that some characteristics of string
theories on pp-waves should be fixed from the requirement of 
modular invariance. 

In the pp-wave case, theta-like function
$\Theta_{(a,b)}(\tau ,\bar{\tau},\nu)$ should appear 
in the closed string partition function. 
It contains a mass parameter $\nu$ and has peculiar properties 
under modular transformations
\begin{eqnarray}
&&\Theta_{(a,b)}(\tau +1,\bar{\tau}+1;\nu)
=\Theta_{(a,b+a)}(\tau ,\bar{\tau};\nu)\,,\quad 
\Theta_{(a,b)}(-\frac{1}{\tau},-\frac{1}{\bar{\tau}};|\tau |\nu)
=\Theta_{(b,-a)}(\tau ,\bar{\tau};\nu)\,.\nonumber
\end{eqnarray}
Notably, the mass parameter $\nu$ changes into
$|\tau |\nu$ under $S$-transformation $\tau \rightarrow -1/\tau$, 
and it gives us severe constraints in 
constructing modular invariant partition functions.
In other words, we can make modular 
invariant partition functions with this clue to go upon.
Now we will study a certain class of 
modular-invariant partition functions on the pp-wave background 
by using the modular properties of 
$\Theta_{(a,b)}(\tau ,\bar{\tau}, \nu)$.

To simplify the problem, we put the following ansatz: \\ (1) There are
several kinds of mass parameters $\nu$'s.\\ (2) For each $\nu$, the
partition function $Z_B(\tau ,\bar{\tau },\nu)$ of the boson and that of
fermion $Z_F(\tau ,\bar{\tau} ,\nu)$ cancel.  That is to say, $Z_B(\tau
,\bar{\tau}, \nu)\cdot Z_F(\tau , \bar{\tau}, \nu)=1 $.\\ We impose the
ansatz (2) because the modular $S$-transformation changes the mass
parameter $\nu$ into another one $|\tau | \nu$ and it is generally
difficult to construct modular invariant combinations of $Z_B$'s and
$Z_F$'s.  In order to avoid this complicated problem, we take the
simplest ansatz $Z_B(\tau ,\bar{\tau}, \nu)\cdot Z_F(\tau , \bar{\tau},
\nu)=1 $ here.  But we should emphasize that there might be other
modular invariant combinations without our ansatz and we cannot say
there are no other possibilities.  Here we will investigate such
restricted cases only and compare our results with the models proposed
earlier.

Now we will classify possible models by the use of the above ansatz.
In order to realize the condition (2), 
the degrees of freedom of bosons must be identical with those 
of fermions. 
When we consider a transverse $D$ dimensional space, 
the degrees of freedom of bosons are $D$ (i.e., $\sharp(\mbox{boson})=D$). 
On the other hand, the degrees of freedom of spinors in $D$ dimensions are
evaluated as
\begin{eqnarray}
\sharp(\mbox{fermion})=2^{[\frac{D}{2}]+\epsilon}\,, 
\quad \epsilon =\left\{
\begin{array}{rcl}
0 & \mbox{Majorana or Weyl}\\
-1 & \mbox{Majorana and Weyl}\\
+1 & \mbox{otherwise}
\end{array}
\right.\nn
\end{eqnarray}
The value of $\epsilon$ depends on what kinds of spinors we consider. 
From the consideration of dimensionality, 
we can understand that the matching of degrees of freedom between 
bosons and fermions happens for $D=1,2,4$ and $8$ only.
For each $D=1,2,4,8$, 
the corresponding degree of freedom is $1,2,4,8$.
That is, we have to consider four kinds of 
sets containing one, two, four, and eight bosons.
Different sets 
are distinguished from mass parameters $\nu$'s.
Due to these mass terms, Lorentz symmetry is broken
into smaller one.
Next let us classify bosonic parts based on the Lorentz symmetry.

We study superstring theories and hence the dimension of 
transverse space should be eight.
For massless cases, the associated Lorentz symmetry 
is $SO(8)$ and there are eight massless bosons. 
However bosons have mass terms in our massive case.
We set $N_a$ as the number of sets 
with $a(=1,2,4,8)$ bosons with the same mass parameter.
Let $\nu_{a,i}$ $(i=1,2,\cdots ,N_a)$ be mass parameters 
for bosons and fermions in the same set.
Due to mass terms, Lorentz symmetry 
is broken down to 
smaller one
\begin{eqnarray}
&&SO(1)^{N_1}\otimes 
SO(2)^{N_2}\otimes 
SO(4)^{N_4}\otimes 
SO(8)^{N_8}\subset SO(8)\,,\nn \\
&&\quad \mbox{with}\qquad N_1+2N_2+4N_4+8N_8=8\,,
\qquad N_1,N_2,N_4,N_8 \in \mathbb{Z}_{\geq 0}\,.\nn
\end{eqnarray}
{}From this constraint, we can classify possible 
combinations:
\begin{eqnarray}
\begin{array}{|cccc|c|}
\hline
N_1 & N_2 & N_4 & N_8 & \mbox{Symmetry}\\
\hline\hline
0 & 0 & 0 & 1 & SO(8)\\
0 & 0 & 2 & 0 & SO(4)\times SO(4)\\
0 & 2 & 1 & 0 & SO(2)\times SO(2)\times SO(4)\\
2 & 1 & 1 & 0 & SO(1)^{\otimes 2}\times SO(2)\times SO(4)\\
8-2\ell & \ell & 0 & 0 & SO(1)^{\otimes (8-2\ell)}\times SO(2)^{\otimes \ell}
\\\hline
\end{array}\nonumber
\end{eqnarray}
with $\ell =0,1,2,3,4$.
Our type IIA model corresponds to the second case 
$(N_1,N_2,N_4,N_8)=(0,0,2,0)$ in the list and symmetry 
is $SO(4)\times SO(4)$. The type IIB string theory on the 
maximally supersymmetric background corresponds to 
$(N_1,N_2,N_4,N_8) = (0,0,0,1)$. 
All of type IIB pp-wave backgrounds with the above-mentioned bosonic 
isometry are found (for example see \cite{Sakaguchi}) 
and we can construct superstring theories on these backgrounds.

Here we turn to partition functions based on our ansatz.
Each set is labelled by the mass parameter $\nu_{a,i}$
$(i=1,2,\cdots ,N_a\,;\,a=1,2,4,8)$.
The associated partition function of boson 
$Z_B(\tau ,\bar{\tau},\nu_{a,i})$ and that 
of fermion $Z_F(\tau ,\bar{\tau}, \nu_{a,i})$
are written in the massive closed string case
\begin{eqnarray}
&&Z_B(\tau ,\bar{\tau},\nu_{a,i})=
\Theta_{(0,0)}(\tau ,\bar{\tau},\nu_{a,i})\,,\quad 
Z_F(\tau ,\bar{\tau},\nu_{a,i})=
\Theta_{(0,0)}(\tau ,\bar{\tau},\nu_{a,i})^{-1}\,.\nn
\end{eqnarray}
Then we can evaluate the total partition function $Z$ as
\begin{eqnarray}
Z=\prod_{a=1,2,4,8}\prod_{i=1}^{N_a}
Z_B(\tau ,\bar{\tau},\nu_{a,i})\cdot Z_F(\tau ,\bar{\tau},\nu_{a,i})
=1\,.\nn
\end{eqnarray}
It is actually modular invariant and 
many models proposed earlier are included in our results.

Last we explain the result $Z=1$ from the point of view of energy
matching.
Let $\varepsilon$ be any 
energy level of states in our string system.
We also introduce $n_B(\varepsilon)$, $n_F(\varepsilon)$
as the number of bosonic states and 
that of fermionic states
at each energy level $\varepsilon$ respectively.
Then the associated partition function $Z$ is 
defined as
\begin{eqnarray}
Z= {\rm Tr} (-1)^{\bf F}e^{-2\pi \tau_2 H}
=\sum_{\varepsilon}(n_B(\varepsilon)-n_F(\varepsilon))
e^{-2\pi \tau_2\varepsilon}\,.\nn
\end{eqnarray}
Here ${\bf F}$ is the fermion number operator and 
we also take $\tau_1 =0$ for simplicity.
By comparing our result $Z=1$, 
we understand following relations
\begin{eqnarray}
&&1=Z=(n_B(\varepsilon =0)-n_F(\varepsilon =0))e^{-2\pi\tau_2 \cdot 0}
+\sum_{\varepsilon >0}
(n_B(\varepsilon )-n_F(\varepsilon ))e^{-2\pi\tau_2 \varepsilon}\,,\nn \\
&&n_B(\varepsilon =0)-n_F(\varepsilon =0)=1\,,\quad 
n_B(\varepsilon )=n_F(\varepsilon )\qquad (\varepsilon >0)\,.\nn
\end{eqnarray}
It shows that 
number of bosonic states and that of fermionic states
match at each energy level $\varepsilon >0$.
Then total energy of bosonic states $E_B=n_B(\varepsilon)\cdot 
\varepsilon$ is equal to 
that of fermionic states $E_F=n_F(\varepsilon)\cdot \varepsilon$
at each energy level $\varepsilon >0$.
On the other hand, there is unbalance in number 
between bosonic states and fermionic states 
for the $\varepsilon =0$ part.
But the associated total energy of bosonic states
$E_B^0=n_B(\varepsilon =0)\cdot 0=0$ 
equals to that of fermionic states $E_F^0=n_F(\varepsilon =0)\cdot 0=0$
in this $\varepsilon =0$ sector.
So the partition function $Z$ is nothing but the Witten index 
$Z=\tr (-1)^{\bf F}$ in our massive case.
This fact is already known in previous papers.
Collecting these considerations, 
we conclude that our ansatz (2) 
is equivalent to a condition (Witten index)$=1$.
When we impose this condition (2) on $Z$, 
the resulting partition function is one and does not 
vanish. It also ensures that total energies of states at each
energy level $\varepsilon$ match between bosonic states and fermionic
states.

Our ansatz satisfies sufficient conditions to construct modular
invariant partition functions.  But we do not know necessary condition
for this problem.  We think it is important to find some further extra
constraints in order to construct consistent string backgrounds and
classify possible strings for massive cases.

\section{Conclusions and Discussions} 

We have discussed the partition function of type IIA string theory
obtained from the eleven-dimensional theory through the
$S^1$-compactification of a transverse direction.

The modular invariance of our type IIA string theory has been proven.
This type IIA string theory is less supersymmetric but it is modular
invariant by virtue of the cancellation between bosonic and fermionic
degrees of freedom.  

We have constructed the boundary states and classified the D-brane
instantons in our theory. The resulting list of the allowed D-brane
instantons is consistent with that of the allowed D-branes obtained
previously in different frameworks.  In addition, we have calculated the
amplitude between D-branes in the closed and open string descriptions,
and checked the channel duality in our theory.  Furthermore, we have
briefly discussed general modular properties. There are many
non-maximally supersymmetric pp-wave backgrounds, but not all of them
would give `modular invariant' superstring theories. Thus, we believe
that the modular invariance is an available clue to classify the `{\it
physical\,}' string theories on pp-waves.

\newpage

%\vspace*{0.5cm}
\noindent 
{\bf\large Acknowledgement}

K.Y thanks Makoto Sakaguchi for useful discussions and comments. 
The work of H.S. was supported by Korea Research Foundation Grant
KRF-2001-015-DP0082. The work of K.S. 
is supported in part by the Grant-in-Aid from the 
Ministry of Education, Science, Sports and Culture of Japan 
($\sharp$ 14740115).

\end{document}